\documentclass[11pt]{article}
\usepackage[margin=1in]{geometry}
\usepackage{amsmath,amssymb,amsthm,mathtools}
\usepackage{booktabs}
\usepackage{array}
\usepackage{enumitem}
\usepackage{microtype}
\usepackage{tikz}
\usepackage[hidelinks]{hyperref}
\usepackage[capitalise,nameinlink]{cleveref}

\newtheorem{theorem}{Theorem}[section]
\newtheorem{proposition}[theorem]{Proposition}
\newtheorem{lemma}[theorem]{Lemma}
\newtheorem{corollary}[theorem]{Corollary}
\theoremstyle{definition}
\newtheorem{definition}[theorem]{Definition}
\newtheorem{example}[theorem]{Example}
\newtheorem{question}[theorem]{Question}
\theoremstyle{remark}
\newtheorem{remark}[theorem]{Remark}
\numberwithin{equation}{section}
\numberwithin{table}{section}
\numberwithin{figure}{section}
\crefname{question}{Question}{Questions}
\crefname{example}{Example}{Examples}

\newcommand{\N}{\mathbb{N}}
\newcommand{\Z}{\mathbb{Z}}
\newcommand{\pair}[2]{\langle #1,\,#2\rangle}
\newcommand{\Gp}[2]{G_{\pair{#1}{#2}}}
\newcommand{\even}{2\N}
\newcommand{\odd}{2\N-1}
\newcommand{\ceil}[1]{\lceil #1\rceil}
\newcommand{\floor}[1]{\lfloor #1\rfloor}
\newcommand{\Ch}{\Gamma}      
\newcommand{\run}{\lambda}    
\newcommand{\Res}{\mathcal{R}}

\title{Up or down at every integer:\\ Collatz-type maps of arbitrary sets and the up-down graph}
\author{Oleg Vasilyev\\
[2pt] {\small California, USA}\\
{\small\texttt{vasilyev.oleg@gmail.com}}}
\date{}

\begin{document}

\maketitle

\begin{abstract}
For a set $S$ of positive integers let $T_S$ send $x$ to $\ceil{3x/2}$ if $x\in S$ and to
$\ceil{x/2}$ otherwise. The odd numbers give the Collatz map, the even numbers give it again after
a shift by one, and $S=\N$ gives Mahler's map $x\mapsto\ceil{3x/2}$. All these maps are subgraphs
of the up-down graph $G$, in which every $x$ has the edges $x\to\ceil{x/2}$ and
$x\to\ceil{3x/2}$, and so are the families of maps in which the direction is forced at some
numbers and free at the others. The cycles of the maps are the simple cycles of the subgraph,
their divergent orbits are its rays, and all maps of a family are bounded exactly when the
subgraph has no ray. We classify the simplest sets and show that density says little: some sets
of density one send every orbit to $1$, and some periodic sets of arbitrarily small density have
increasing orbits. Every positive integer is the least element of a cycle of some map. The critical
proportion of steps up is $\log_32$: cycles stay below it, divergent orbits stay above it, and both
come arbitrarily close. For sets periodic modulo a power of $2$, an effective up-frequency from a
finite Markov chain takes the place of the density, and a finite computation can prove that every
orbit reaches a cycle. Mahler's $Z$-number problem is implied by a question about one such map.
Finally, we ask how far the Collatz map is from maps whose behaviour can be proved. Finite changes
never help. Adding fewer than $\log_{3/2}N+1$ even numbers below $N$ makes an orbit increase, while
removing the odd numbers congruent to $9$ or $15$ modulo $16$, an eighth of all integers, gives a
map all of whose orbits provably reach $(1,2)$; removing $22$ residue classes modulo $512$, a
proportion $0.043$, also suffices.
\end{abstract}

\medskip
\noindent\textbf{Keywords:} Collatz problem; generalized Collatz maps; directed
graphs; Markov chains; Mahler's $Z$-numbers; $2$-adic integers.

\noindent\textbf{2020 Mathematics Subject Classification:} Primary 11B83; Secondary 05C20, 11J71, 11Y55, 60J10.


\section{Introduction}\label{sec:intro}

\subsection{Up or down at every integer}

The Collatz map
\[
T(x)=\begin{cases}(3x+1)/2, & x \text{ odd},\\ x/2, & x\text{ even},\end{cases}
\]
changes a positive integer by about half of its value at every step: odd numbers go up and even
numbers go down. The Collatz conjecture states that every orbit reaches the cycle $(1,2)$; see
the surveys and bibliographies of Lagarias \cite{Lag85,Lag10,LagBib1,LagBib2}. In \cite{Vas26} we
kept the size of the steps and let the $2$-adic valuation of the current number decide their
direction. Here the direction is decided at every integer separately, by an arbitrary set. Let $\N=\{1,2,3,\dots\}$.

\begin{definition}\label{def:TS}
For $x\in\N$ let $U(x)=\ceil{3x/2}$ and $D(x)=\ceil{x/2}$, and for $S\subseteq\N$ let
$T_S\colon\N\to\N$ be
\begin{equation}\label{eq:TS}
T_S(x)=\begin{cases}U(x), & x\in S,\\ D(x), & x\notin S.\end{cases}
\end{equation}
\end{definition}

So $U(x)=(3x+1)/2$ and $D(x)=(x+1)/2$ for odd $x$, and $U(x)=3x/2$ and $D(x)=x/2$ for even
$x$. A step up adds $\ceil{x/2}$ to $x$, a step down subtracts $\floor{x/2}$, and a step down at an
even number halves it only once, rather than removing all factors $2$ at once. These are exactly
the two steps of the Collatz map, together with the step $x\mapsto(x+1)/2$ at odd numbers that was
used by the maps $T^-_S$ of \cite{Vas26}. Since $U(x)>x$ for all $x$ and $D(x)<x$ for all
$x\ge2$, while $D(1)=1$, an orbit of $T_S$ goes up exactly at the elements of $S$ that it visits.

\begin{remark}[Notation]
We keep the letter $T$, which is customary for the Collatz map: with our notation the Collatz map
$T$ is $T_{\odd}$. Readers of \cite{Vas26} should note that the symbol $T_S$ is used differently
there: in \cite{Vas26} the set $S$ is a set of exponents of $2$, the direction of a step depends
on the exponent of $2$ in the current number, and a step down at an even number removes all
factors $2$ at once. In this paper $S$ is a set of numbers, the direction depends only on whether
the current number lies in $S$, and every step divides by $2$ once. Only four maps belong to
both families: $T_\N$, $T^-_\N$, $T_{[2,\infty)}$ and $T^-_{[2,\infty)}$ of \cite{Vas26} are the
maps $T_\N$, $T_{\even}$, $T_{\N\setminus(4\N-2)}$ and $T_{4\N}$ here.
\end{remark}

The family runs from $T_\emptyset=D$, under which every orbit decreases to the fixed point $1$, to
$T_\N=U$, under which every orbit increases. The orbits of $U$ are those of Mahler's $3/2$ problem
\cite{Mah68}; for example, $U^r(1)=\floor{K(3/2)^r}$ with $K=1.6222705\ldots$ \cite{OW91}. The
family contains the Collatz map twice:
\[
T_{\odd}=T,\qquad T_{\even}(y+1)=T(y)+1\quad(y\ge0),
\]
where $T(0)=0$. Indeed, for odd $y$ the number $y+1$ is even, and $U(y+1)=\frac{3y+3}{2}=T(y)+1$,
while for even $y$ we get $D(y+1)=\frac{y+2}{2}=T(y)+1$. So $T_{\even}$ is the Collatz map on
$\{0,1,2,\dots\}$ shifted by one, with the fixed point $1$ and the cycle $(2,3)$ in place of $0$
and $(1,2)$; it is the map $T^-_\N$ of \cite[Proposition 6.1]{Vas26}. For example,
\begin{align*}
T_{\even}&:\quad 8,\,12,\,18,\,27,\,14,\,21,\,11,\,6,\,9,\,5,\,3,\,2,\,3,\dots\\
T_{\{\text{primes}\}}&:\quad 7,\,11,\,17,\,26,\,13,\,20,\,10,\,5,\,8,\,4,\,2,\,3,\,5,\dots\\
T_{\{\text{squares}\}}&:\quad 49,\,74,\,37,\,19,\,10,\,5,\,3,\,2,\,1,\,2,\,1,\dots
\end{align*}
The first orbit is the Collatz orbit of $7$ shifted by one and ends in $(2,3)$; the second ends
in the cycle $(2,3,5,8,4)$ and the third in $(1,2)$.

An orbit of a map $f\colon\N\to\N$ is either eventually periodic, and then ends in a
\emph{cycle}, which we write as a tuple that starts with its least element, or it consists of
distinct numbers; then it tends to infinity, and we say that it \emph{diverges}. We call $T_S$
\emph{bounded} if all its orbits are eventually periodic, and \emph{unbounded} otherwise. If $S$ is
finite, $T_S$ is bounded: it maps $[1,U(\max S)]$ into itself, and every larger number goes down.
The questions of this paper are:
\begin{enumerate}[label=(\arabic*)]
\item Which sets $S$ give bounded maps, and which cycles do the maps have?
\item How are the maps of the family related to each other?
\item Where, between the bounded and the unbounded maps, does the Collatz map lie, and how far is
it from maps whose behaviour can actually be proved?
\end{enumerate}
The rest of this introduction describes the answers we found.

\subsection{Maps as subgraphs of one graph}

All maps $T_S$ are built from the same two steps, so they all reside in one graph. The
\emph{up-down graph} $G$ has the vertex set $\N$ and, at every vertex $x$, the two edges
$x\to U(x)$ and $x\to D(x)$ (\cref{def:G}, \cref{fig:G}). A map $T_S$ is the subgraph $G_S$ that
keeps one of the two edges at every vertex, the up-edge at the elements of $S$ and the down-edge
elsewhere. A \emph{family} of maps, in which the step up is forced on a set $A$, the step down is
forced outside a set $B\supseteq A$, and both are allowed in between, is the subgraph $\Gp AB$ that
keeps one or two edges at every vertex, and the family of all maps is $G$ itself.

This gives a correspondence between maps and subgraphs (\cref{prop:dict}): the cycles of the maps of a
family are exactly the simple cycles of its subgraph, any set of pairwise disjoint simple cycles
occurs in a single map, the divergent orbits are exactly the rays (infinite simple paths), and all
maps of the family are bounded if and only if the subgraph has no ray. So a classification of maps
is the same as a classification of subgraphs of $G$, and statements about all maps at once are
statements about $G$. Boundedness depends only on $S$ ``at infinity'': changing $S$ at finitely
many numbers does not change whether the map has a divergent orbit (\cref{lem:finite}).

\subsection{The simplest classification}

The first distinction is between sets that obviously give bounded maps and the others. If $S$ has
no element in an interval $(2B/3,B]$, then no step leads from $[1,B]$ above $B$, so $T_S$ maps
$[1,B]$ into itself; we call such a $B$ a \emph{barrier}. A set with infinitely many barriers is
\emph{trivially bounded}: every orbit is trapped below some barrier. This happens exactly when $S$
is finite or, writing $S=\{s_1<s_2<\cdots\}$, when infinitely often the next element $s_{i+1}$ is
out of reach of a step up from $s_i$, that is, $s_{i+1}>U(s_i)$ (\cref{prop:trivial}). Every
other set has at least $\log_{3/2}N-O(1)$ elements below $N$, and this is sharp.

For the remaining sets the important structure is that of the step up alone. The map $U$ is
injective, and $\N$ splits into the \emph{chains} $\{c,U(c),U^2(c),\dots\}$, one for each
$c\equiv1\pmod 3$; the chains are the orbits of $T_\N$. An orbit of $T_S$ makes several steps up
in a row exactly when $S$ contains several consecutive elements of a chain; in particular, it goes
up at every step (we call such an orbit \emph{increasing}) exactly when $S$ contains all elements
of a chain from the starting number on. If no element of $S$
goes up to another element of $S$, every number above $3$ falls below itself within two steps, and
$T_S$ is bounded with one or two of the cycles $(1)$, $(1,2)$, $(2,3)$, $(3,5)$; this holds for
the squares and for a set of density $3/5$. Two steps up in a row already allow divergent orbits
(\cref{prop:runs}).

The density of $S$ turns out to say very little about $T_S$. A single chain has density zero
and gives an increasing orbit; even periodic sets of arbitrarily small density can give
increasing orbits (\cref{prop:depth}). In the other direction, if the complement of $S$ is closed
under the step down, then an orbit that leaves $S$ never returns to it and falls to $1$. This
``one-way door'' gives sets of density one whose maps send every orbit to the fixed point $1$
(\cref{thm:density1}). \Cref{tab:classes} in \cref{sec:classification} collects these classes.

\subsection{Cycles across all maps}

By definition, the cycles of all maps together are the simple cycles of $G$. There are
infinitely many of them: $G$ is strongly connected, and every positive integer is the least element
of a cycle of some map, even of a cycle with a single run of steps up (\cref{thm:G}). On the
other hand there are only finitely many cycles of each length, because a cycle of length $\ell$
with $a$ steps up satisfies $3^a<2^\ell$ with a gap controlled by its least element. Any set of
disjoint cycles is the set of cycles of a single map (\cref{prop:prescribe}), so the cycles of
single maps carry no further restriction. \Cref{tab:census} lists all cycles of length at most
$30$; every one of them has $a=\floor{\ell\log_32}$, and we ask whether this always holds.

\subsection{Near the threshold}

Which sets are, intuitively, the most expanding ones whose maps can still be proved to be
bounded, and which are the least expanding ones whose maps can be proved to be unbounded? Since
density is not the right yardstick, we measure how often an orbit goes up. A step up multiplies by
about $3/2$ and a step down by about $1/2$, so the critical proportion of steps up is
$\log_32=0.6309\ldots$, the solution of $(3/2)^p(1/2)^{1-p}=1$. It is sharp from both sides
(\cref{sec:threshold}): every cycle has a proportion $a/\ell<\log_32$ of steps up, and these
proportions come arbitrarily close to $\log_32$; every divergent orbit has a proportion of steps
up that is asymptotically at least $\log_32$, and there are divergent orbits whose proportion is
arbitrarily close to $\log_32$ (\cref{thm:barely}).

For sets that are periodic modulo a power of $2$ the proportion of steps up of a typical orbit can
be computed. Following an orbit modulo $2^k$, each step reveals one new binary digit of the
starting number, and the residues form a Markov chain with transition probabilities $1/2$
(\cref{lem:fresh}). Its stationary distribution gives an \emph{effective up-frequency} $\pi$, which
is $1/2$ for the Collatz map. It is $\pi$, not the density of $S$, that decides on which side of
the threshold a typical orbit lies: if $\pi<\log_32$, then almost every number eventually falls
below itself, as Terras and Everett proved for the Collatz map \cite{Ter76,Eve77}
(\cref{thm:terras}), and $\pi$ and the density can lie on opposite sides of $\log_32$
(\cref{tab:pi}). A stronger, uniform version of descent can be decided by a finite computation on
the residue graph, and when it holds it yields a complete proof that every orbit reaches one of
finitely many cycles, with a certificate that can be checked by hand (\cref{thm:uniform}). Finally,
Mahler's question whether $Z$-numbers exist \cite{Mah68} would be answered by a statement about one
periodic map of the family, with $\pi=5/6$ (\cref{prop:mahler}).

\subsection{How far is the Collatz map from provable behaviour?}

The Collatz map is $T_S$ for the odd numbers. Which modifications of this set make the behaviour of
the map provable, either that all orbits reach $(1,2)$ or that some orbit diverges, and how small
can such a modification be? \Cref{tab:distance} summarizes the answers of \cref{sec:collatz}.

\begin{table}[ht]
\centering
\caption{Modifications of the Collatz set $\odd$ and what they make provable (\cref{sec:collatz}).
The middle column counts the modified numbers below $N$.}\label{tab:distance}
\begin{tabular}{@{}>{\raggedright\arraybackslash}p{0.43\textwidth}>{\raggedright\arraybackslash}p{0.2\textwidth}>{\raggedright\arraybackslash}p{0.3\textwidth}@{}}
\toprule
modification of $\odd$ & size below $N$ & provable behaviour\\
\midrule
any finite change & $O(1)$ & none: divergence is unaffected (\cref{cor:finite})\\
add the even numbers $2,8,12,18,62,140,210,710,\dots$ of the chain of $1$ & $<\log_{3/2}N+1$ &
the orbit of $1$ increases (\cref{prop:mahleradd})\\
remove the odd numbers in sparse intervals $(2B_i/3,B_i]$ & lower density $0$, upper density
$\ge1/6$ & bounded (\cref{prop:barrierremove})\\
remove the odd $x\equiv9,15\pmod{16}$ & $N/8+O(1)$ & every orbit reaches $(1,2)$
(\cref{thm:mod16})\\
remove $22$ odd residue classes modulo $512$ & $0.043N+O(1)$ & every orbit reaches $(1,2)$
(\cref{tab:deltak})\\
\bottomrule
\end{tabular}
\end{table}

So a logarithmically thin addition already gives a provably divergent orbit, while the smallest
periodic removal that we found to give provable convergence changes $4.3\%$ of all integers. Its
proof is a finite certificate of the same kind as above, and finding the smallest such removal
modulo $2^k$ is a hitting-set problem for the cycles of the residue graph that expand, which are
the shadows of the cycles of the Collatz map at negative rational numbers, such as $-1$ and
$-5\to-7\to-10\to-5$ (\cref{rem:shadows}). We do not know whether the proportion of changed
numbers can be made arbitrarily small, or whether removing odd numbers alone can ever produce a
divergent orbit.

\subsection*{Related work}
The map $T_\N=U$ appears in Mahler's work on $Z$-numbers \cite{Mah68}, its orbit of $1$ in
Odlyzko and Wilf's study of the Josephus problem \cite{OW91}, the fractional parts of $\xi(3/2)^n$
in the work of Flatto, Lagarias and Pollington \cite{FLP95}, and related questions in the number
systems with rational base of Akiyama, Frougny and Sakarovitch \cite{AFS08}. Mignosi \cite{Mig95}
replaced the multiplier $3/2$ at odd numbers by an arbitrary $\beta>1$. Steiner \cite{Ste78} and
Simons and de Weger \cite{SdW05} bounded the Collatz cycles with few runs of odd steps, and the
fact that almost all orbits of the Collatz map fall below their starting value is due to Terras
\cite{Ter76} and Everett \cite{Eve77}. Applegate and Lagarias \cite{AL06} proved Farkas's weak
$3x+1$ conjecture, a statement that steps combined freely reach every positive integer;
\cref{thm:G}(a) is a statement of the same kind for $G$. Generalized Collatz maps
\cite{Con72,MW84,KS07} contain $T_S$ for every periodic $S$, and questions about them are
undecidable in general \cite{Con72,KS07}. The counting argument behind our \cref{lem:fresh},
that every step reveals one new binary digit, goes back to Terras \cite{Ter76}; see
\cite{Lag85}. Yolcu, Aaronson and Heule \cite{YAH23} encoded the Collatz map and some of its
variants as string rewriting systems and proved termination for several variants automatically;
the weight certificates of \cref{sec:uniform,sec:collatz} are proofs of a much simpler kind.
Monks, Monks, Monks and Monks \cite{MMMM13} used the cycles of the residue graphs of the Collatz map to show that every divergent Collatz orbit must meet certain residue classes, among them the classes $9,15\pmod{16}$ of \cref{thm:mod16}; see \cref{rem:sufficient}.
Choices between steps appear in a two-player game of Alth\"ofer, Hartisch and Zipproth \cite{AHZ23} and in the branching maps of \cite[Section 7]{Vas26}. We have not found the maps $T_S$ for arbitrary $S$, or the graph $G$, in the literature surveyed in \cite{LagBib1,LagBib2}.

\subsection*{Organization and notation}
\Cref{sec:graph} introduces the graph $G$ and the dictionary, \cref{sec:classification} the
simplest classes of sets, and \cref{sec:cycles} the cycles of all maps. \Cref{sec:threshold} is
about the threshold $\log_32$ and periodic sets, \cref{sec:collatz} about modifications of the
Collatz set, and \cref{sec:questions} collects open questions. \Cref{sec:computations} describes
the computations.

For integers $a\le b$ we write $[a,b]=\{a,\dots,b\}$ and $[a,\infty)=\{a,a+1,\dots\}$; for real
numbers $u<v$, $(u,v]$ denotes the set of integers $x$ with $u<x\le v$. We write $\#A$ for the
number of elements of a finite set $A$, $\N_0=\{0,1,2,\dots\}$, $\even$ and $\odd$ for the even
and the odd positive integers, and similarly $4\N=\{4,8,\dots\}$, $4\N-2=\{2,6,10,\dots\}$,
$4\N-3=\{1,5,9,\dots\}$ and $4\N-1=\{3,7,11,\dots\}$. A set $S$ has (natural) density $\delta$ if
$\#(S\cap[1,N])/N\to\delta$, and lower and upper density are the $\liminf$ and the $\limsup$ of
this ratio. We write $\{t\}=t-\floor t$ for the fractional part of a real number $t$, and $x\bmod m$
for the least nonnegative residue. A set $S$ is \emph{periodic modulo $m$} if whether $x\in S$
depends only on $x\bmod m$.

\section{The maps and the up-down graph}\label{sec:graph}

\subsection{The two steps}

We collect the elementary properties of the two steps that are used throughout. The only
non-obvious one is (d): iterating the step up behaves like multiplying by $3/2$, up to a bounded
error that is fixed once and for all by the starting number.

\begin{lemma}\label{lem:steps}
Let $x,y\in\N$ and $e=x\bmod2$.
\begin{enumerate}[label=(\alph*)]
\item $U(x)=(3x+e)/2$ and $D(x)=(x+e)/2$. The map $U$ is strictly increasing and $D$ is
non-decreasing; $U(x)>x$, and $D(x)<x$ for $x\ge2$.
\item $D^s(y)=\ceil{y/2^s}$ for $s\ge0$. In particular every orbit of $D$ reaches $1$.
\item For an integer $B$, $U(y)>B$ if and only if $y>2B/3$.
\item $3y/2\le U(y)$ and $U(y)+1\le\frac32(y+1)$. Hence for every $x$ there is a number
$K_x\in(x,x+1)$ with $K_x(3/2)^r-1<U^r(x)<K_x(3/2)^r$, that is, $U^r(x)=\floor{K_x(3/2)^r}$, for
all $r\ge0$.
\item The image of $U$ is $\{y\ge2:y\not\equiv1\pmod3\}$.
\item $U(x+2m)=U(x)+3m$ and $D(x+2m)=D(x)+m$ for all $m\in\N_0$.
\end{enumerate}
\end{lemma}

\begin{proof}
(a) is immediate. (b) follows from $\ceil{\ceil{y/a}/b}=\ceil{y/(ab)}$ for integers $y$ and
$a,b\ge1$. (c) $\ceil{3y/2}\ge B+1$ if and only if $3y/2>B$. (d) The first inequality is clear,
and $U(y)+1\le(3y+1)/2+1=\frac32(y+1)$. So $a_r=(2/3)^rU^r(x)$ is non-decreasing in $r$,
$b_r=(2/3)^r(U^r(x)+1)$ is non-increasing, and $b_r-a_r=(2/3)^r$ tends to $0$; both converge to
some $K_x$ with $a_r\le K_x\le b_r$. If $a_r=K_x$ for some $r$, then $a_{r'}=a_r$ for all $r'\ge r$,
so $U^{r'}(x)=(3/2)^{r'-r}U^r(x)$ would be an integer for all $r'\ge r$, which is impossible.
Similarly $b_r=K_x$ would make $(3/2)^{r'-r}(U^r(x)+1)$ an integer for all $r'\ge r$. So
$a_r<K_x<b_r$ for all $r$, which is the claim; for $r=0$ it gives $x<K_x<x+1$. (e)
$U(2k)=3k$ and $U(2k-1)=3k-1$. (f) $x$ and $x+2m$ have the same parity.
\end{proof}

For example, $K_1=1.6222705\ldots$ is the constant of Odlyzko and Wilf \cite{OW91}, and
$U^r(1)=\floor{K_1(3/2)^r}$ for all $r$.

\subsection{The up-down graph}

\begin{definition}\label{def:G}
The \emph{up-down graph} $G$ is the directed graph with vertex set $\N$ and, at every vertex
$x$, an \emph{up-edge} $x\to U(x)$ and a \emph{down-edge} $x\to D(x)$ (a loop at $x=1$). For
$A\subseteq B\subseteq\N$ let $\Gp AB$ be the spanning subgraph that keeps the up-edge at $x$ if
and only if $x\in B$ and the down-edge if and only if $x\notin A$, and let $G_S=\Gp SS$. We also
write $\pair AB$ for the family of the maps $T_S$ with $A\subseteq S\subseteq B$.
\end{definition}

\begin{figure}[ht]
\centering
\begin{tikzpicture}[x=1.1cm,>=stealth,
  vtx/.style={circle,draw,inner sep=0pt,minimum size=17pt,font=\small},
  upe/.style={->,thick},
  dne/.style={->,densely dashed,black!65}]
\foreach \i in {1,...,12} \node[vtx] (v\i) at (\i,0) {\i};
\foreach \a/\b/\bnd in {1/2/60,2/3/60,3/5/50,4/6/50,5/8/42,6/9/42,7/11/36,8/12/36}
  \draw[upe] (v\a) to[bend left=\bnd] (v\b);
\foreach \a/\b/\bnd in {2/1/60,3/2/60,4/2/50,5/3/50,6/3/42,7/4/42,8/4/36,9/5/36,10/5/32,11/6/32,12/6/30}
  \draw[dne] (v\a) to[bend left=\bnd] (v\b);
\draw[dne] (v1) to[out=235,in=305,looseness=8] (v1);
\foreach \a/\b in {9/14,10/15,11/17,12/18}
  \draw[upe] (v\a) -- ++(0.3,0.75) node[above,font=\scriptsize,inner sep=1pt] {\b};
\end{tikzpicture}
\caption{The up-down graph $G$ on $[1,12]$. Solid arcs above the line are up-edges
$x\to\ceil{3x/2}$ (those leaving $[1,12]$ are short arrows marked with their targets), dashed arcs
below the line are down-edges $x\to\ceil{x/2}$, and the loop is the down-edge at $1$. A map $T_S$
keeps the solid edge at the elements of $S$ and the dashed edge elsewhere.}\label{fig:G}
\end{figure}
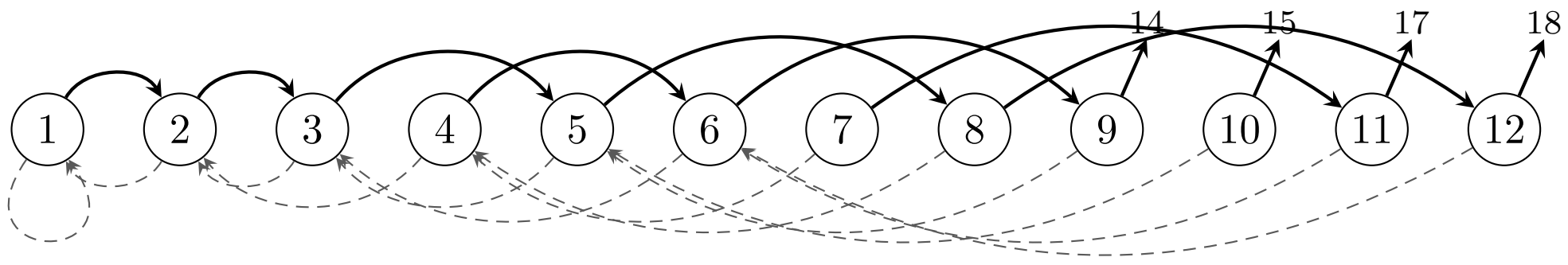

For example, $\Gp\emptyset{\even}$ keeps all down-edges and the up-edges at the even numbers; it
is the union of the graphs of all maps $T_S$ with $S\subseteq\even$, that is, of all maps obtained
from the shifted Collatz map $T_{\even}$ by turning some of its steps up into steps down. Similarly
$\Gp{\odd}\N$ belongs to the maps that go up at every odd number and are free at the even numbers.

Thus $G_S$ is the graph of $T_S$, with one out-edge at every vertex, and
$\Gp AB=\bigcup_{A\subseteq S\subseteq B}G_S$: on $A$ the step up is forced, outside $B$ the step
down is forced, and on $B\setminus A$ both steps are allowed. Every spanning subgraph of $G$ in
which each vertex keeps at least one of its two out-edges is $\Gp AB$ for exactly one pair
$A\subseteq B$, and $G=\Gp\emptyset\N$. So
\begin{itemize}
\item the sets $S$ correspond to the subgraphs with all out-degrees $1$,
\item the families $\pair AB$ correspond to the subgraphs with all out-degrees at least $1$, and
\item the family of all maps corresponds to $G$.
\end{itemize}
As in \cite[Lemma 7.2]{Vas26}, any collection $\mathcal S$ of sets gives the same subgraph
$\bigcup_{S\in\mathcal S}G_S$ as the family $\pair{\bigcap\mathcal S}{\bigcup\mathcal S}$, so the
branching families of \cite[Section 7]{Vas26} correspond to these subgraphs as well.

Two subgraphs are worth keeping in mind. The edges of $G_\emptyset$ lead from $2y-1$ and $2y$
to $y$, so $G_\emptyset$ is a binary tree with root $1$ (and a loop at $1$); in binary, the step
down sends $x-1$ to $\floor{(x-1)/2}$, that is, it deletes the last binary digit of $x-1$. The
graph $G_\N$ is a disjoint union of rays, the chains of \cref{sec:runs}. A path is \emph{simple}
if its vertices are distinct, a \emph{ray} is an infinite simple path, and a \emph{simple cycle} is
a closed path whose vertices are distinct.

\subsection{The dictionary}

\begin{proposition}[Dictionary]\label{prop:dict}
Let $A\subseteq B\subseteq\N$ and $H=\Gp AB$.
\begin{enumerate}[label=(\alph*)]
\item The cycles of the maps in $\pair AB$ are exactly the simple cycles of $H$. A set of simple
cycles of $H$ consists of cycles of a single map in $\pair AB$ if and only if the cycles are
pairwise vertex-disjoint.
\item The divergent orbits of the maps in $\pair AB$ are exactly the rays of $H$.
\item All maps in $\pair AB$ are bounded if and only if $H$ has no ray, if and only if from every
$x$ only finitely many vertices can be reached in $H$.
\end{enumerate}
\end{proposition}

The point of the proof is that a simple cycle or a ray visits every vertex at most once, so that
the choices it makes between the two edges never conflict and can be made by a single set $S$.
For example, in \cref{fig:G} the closed path $2\to3\to5\to8\to4\to2$ is a simple cycle of $G$: it
uses the up-edges at $2$, $3$, $5$ and the down-edges at $8$ and $4$. So $(2,3,5,8,4)$ is a cycle
of $T_S$ for every set $S$ that contains $2,3,5$ and does not contain $4,8$, such as the set of
primes, and of no other map. By contrast, the closed path $2\to3\to2\to1\to2$ is not simple: it
would need both edges at $2$, and no single map follows it.

\begin{proof}
(a) A cycle of $T_S$ uses edges of $G_S\subseteq H$ and visits distinct numbers. Conversely, let
$\mathcal C$ be a set of pairwise disjoint simple cycles of $H$, let $V$ be the union of their
vertex sets, and let $V^\uparrow\subseteq V$ be the set of vertices at which they use the
up-edge. Put $S=(A\setminus V)\cup V^\uparrow$. Every edge of a cycle is an edge of $H$, so
$V^\uparrow\subseteq B$, and a vertex of $V\cap A$ has no down-edge in $H$ and lies in
$V^\uparrow$. Hence $A\subseteq S\subseteq B$, and $T_S$ follows every cycle of $\mathcal C$.
Two cycles of one map are equal or disjoint, since $G_S$ has one out-edge at every vertex.
(b) A divergent orbit consists of distinct numbers, so it is a ray of $G_S\subseteq H$.
Conversely, for a ray of $H$ the construction of (a), with $V$ the vertex set of the ray, gives
$S$ with $A\subseteq S\subseteq B$ such that the orbit of the first vertex follows the ray; this
orbit consists of distinct numbers, so it diverges. (c) The first equivalence follows from (b).
If infinitely many vertices can be reached from $x$, the tree in which every reachable vertex
other than $x$ is attached to the vertex from which a breadth-first search first reaches it is
infinite and has finite degrees, so by K\"onig's lemma it has an infinite path from $x$, which is a
ray of $H$; compare \cite[Lemma 7.3]{Vas26}. Conversely, a ray from $x$ reaches infinitely many
vertices.
\end{proof}

In \cite[Proposition 7.4]{Vas26} boundedness passed from a family to its members but not back,
because there a single exponent decided the direction at infinitely many numbers; here (c) is an
equivalence. \Cref{tab:dict} summarizes the correspondence. It is a duality between classes of
maps and classes of subgraphs of $G$: statements about all maps at once are statements about $G$,
and statements about all maps of a family are statements about a subgraph of $G$.

\begin{table}[ht]
\centering
\caption{Maps and subgraphs of the up-down graph $G$ (\cref{prop:dict}).}\label{tab:dict}
\begin{tabular}{@{}ll@{}}
\toprule
maps & subgraphs of $G$\\
\midrule
a map $T_S$ & $G_S$: one out-edge at every vertex\\
a family $\pair AB$ & $\Gp AB$: one or two out-edges at every vertex\\
all maps & $G$\\
the orbit of $x$ under $T_S$ & the path from $x$ in $G_S$\\
a cycle of a map in $\pair AB$ & a simple cycle of $\Gp AB$\\
the set of cycles of one map & a set of pairwise vertex-disjoint simple cycles\\
a divergent orbit of a map in $\pair AB$ & a ray of $\Gp AB$\\
all maps in $\pair AB$ bounded & no ray in $\Gp AB$; all reachable sets finite\\
$T_S$ maps $[1,B]$ into itself & no edge of $G_S$ from $[1,B]$ to $(B,\infty)$\\
\bottomrule
\end{tabular}
\end{table}

Two kinds of families will appear in questions. We call a set $B$ \emph{force-bounded} if it
forces boundedness on every map that goes up only inside $B$: $T_S$ is bounded for every
$S\subseteq B$, that is, $\Gp\emptyset B$ has no ray. Dually, we call a set $A$
\emph{force-divergent} if it forces a divergent orbit on every map that goes up on $A$: $T_S$ is
unbounded for every $S\supseteq A$, whatever happens outside $A$. Every subset of a force-bounded
set is force-bounded, and every set that contains a force-divergent set is force-divergent. For
example, a set $A$ that contains all the numbers
$x,U(x),U^2(x),\dots$ for some $x$ is force-divergent, since then the orbit of $x$ is
$x,U(x),U^2(x),\dots$ under every $T_S$ with $S\supseteq A$.

\subsection{Finite changes}

The next lemma says that boundedness is a property of $S$ at infinity. It is simple, but it is the
first answer to the question of \cref{sec:collatz}: no finite modification of the Collatz set can
decide whether the Collatz map has divergent orbits.

\begin{lemma}\label{lem:finite}
Let $S,S'\subseteq\N$ agree on $(m,\infty)$ for some $m$.
\begin{enumerate}[label=(\alph*)]
\item Every divergent orbit of $T_S$ is, from some point on, an orbit of $T_{S'}$. In
particular, $T_S$ is bounded if and only if $T_{S'}$ is bounded.
\item The cycles of $T_S$ that lie in $(m,\infty)$ are cycles of $T_{S'}$.
\end{enumerate}
\end{lemma}

\begin{proof}
(a) A divergent orbit $x_0,x_1,\dots$ tends to infinity, so $x_t>m$ for all $t\ge t_0$, and then
$x_{t+1}=T_S(x_t)=T_{S'}(x_t)$ for $t\ge t_0$. So $x_{t_0},x_{t_0+1},\dots$ is a divergent orbit of
$T_{S'}$. The statement is symmetric in $S$ and $S'$. (b) $T_S$ and $T_{S'}$ agree on
$(m,\infty)$.
\end{proof}

Finite changes can create or destroy cycles below $m$, and so they can change whether
all orbits reach a given cycle; but they cannot make a bounded map out of an unbounded one, or the
other way round.

\section{The simplest classification}\label{sec:classification}

\subsection{Barriers and trivially bounded sets}

Consider an infinite set $S=\{s_1<s_2<\cdots\}$. If the next element $s_{i+1}$ is out of reach of
a step up from $s_i$, that is, if $s_{i+1}>U(s_i)$, then $T_S$ maps $[1,s_{i+1}-1]$ into itself:
the elements of $S$ in this interval are at most $s_i$ and go up to at most $U(s_i)$, and all
other numbers go down. This means that every orbit that starts below $s_{i+1}$ stays below $s_{i+1}$ forever.
The following definition captures exactly this situation. It is based on a simple fact
(\cref{lem:steps}(c)): the numbers $x\le B$ whose step up leads above $B$ are exactly the numbers
in the interval $(2B/3,B]$. If none of them is in $S$, nothing can cross $B$ from below.

\begin{definition}\label{def:barrier}
A \emph{barrier} of $S\subseteq\N$ is an integer $B\ge1$ such that $S$ has no element in the
interval $(2B/3,B]$. The set $S$ is \emph{trivially bounded} if it has infinitely many barriers.
\end{definition}

For example, let $S=\{1,2,4,8,16,\dots\}$ be the set of powers of $2$. Then $B=6$ is a barrier,
since the interval $(4,6]=\{5,6\}$ contains no power of $2$; indeed the only steps up from
$[1,6]$ are $1\to2$, $2\to3$ and $4\to6$, and none of them leaves $[1,6]$. In the same way
$B=3\cdot2^{k-1}$ is a barrier for every $k\ge1$, because the largest power of $2$ below it,
$2^k$, goes up exactly to $3\cdot2^{k-1}$. So the powers of $2$ are trivially bounded: every orbit
of their map is trapped below one of these barriers.

\begin{proposition}\label{prop:trivial}
Let $S\subseteq\N$.
\begin{enumerate}[label=(\alph*)]
\item $B$ is a barrier of $S$ if and only if $T_S$ maps $[1,B]$ into itself, if and only if $G_S$
has no edge from $[1,B]$ to $(B,\infty)$. In $G$ these edges are the up-edges at the vertices in
$(2B/3,B]$.
\item $S$ is trivially bounded if and only if $S$ is finite or $s_{i+1}>U(s_i)$ for infinitely
many $i$. The barriers $B\ge s_1$ are the integers with $U(s_i)\le B<s_{i+1}$ for some $i$ (where
$s_{i+1}=\infty$ if $s_i$ is the largest element of $S$).
\item If $S$ is trivially bounded, then so is every subset of $S$, and $T_S$ is bounded. Thus
trivially bounded sets are force-bounded.
\item If $S$ is not trivially bounded, then $\#(S\cap[1,N])\ge\log_{3/2}N-C$ for some
$C=C(S)$ and all $N\ge1$.
\end{enumerate}
\end{proposition}

\begin{proof}
(a) Let $x\le B$. Then $D(x)\le B$, and $U(x)>B$ if and only if $x>2B/3$ by
\cref{lem:steps}(c). So some $x\le B$ has $T_S(x)>B$ if and only if $S$ meets $(2B/3,B]$.

(b) Every $B<s_1$ is a barrier. For $B\ge s_1$ let $s_i=\max(S\cap[1,B])$, so that
$s_i\le B<s_{i+1}$. Then $B$ is a barrier if and only if $s_i\le2B/3$, that is, if and only if
$U(s_i)\le B$ by \cref{lem:steps}(c). This proves the second statement. If $S$ is finite, every
$B\ge U(\max S)$ is a barrier. If $S$ is infinite, the barriers $B\ge s_1$ are the integers in the
finite intervals $[U(s_i),s_{i+1}-1]$, and there are infinitely many of them if and only if
infinitely many of these intervals are nonempty, that is, $s_{i+1}>U(s_i)$ for infinitely many
$i$.

(c) A barrier of $S$ is a barrier of every subset. Every $x$ lies below some barrier $B$, and by
(a) its orbit stays in $[1,B]$.

(d) The idea is that a set without large barriers must have an element in every interval
$(2B/3,B]$ with $B$ large, and about $\log_{3/2}N$ such intervals fit disjointly below $N$. Let
$B_0$ be larger than all barriers of $S$, and put $B_{k+1}=U(B_k)$. Since
$2B_{k+1}/3\ge B_k$, the intervals $(2B_k/3,B_k]$ are pairwise disjoint, and each contains an
element of $S$, because $B_k$ is not a barrier; so $\#(S\cap[1,B_k])\ge k$. By \cref{lem:steps}(d), $B_k+1\le(3/2)^k(B_0+1)$.
For $B_k\le N<B_{k+1}$ this gives
$\#(S\cap[1,N])\ge k\ge\log_{3/2}\bigl((N+1)/(B_0+1)\bigr)-1$.
\end{proof}

For a trivially bounded $S$ with barriers $B_1<B_2<\cdots$, the map $T_S$ restricts to self-maps
of the nested finite sets $[1,B_k]$, and all questions about $T_S$ are finite questions. Finite
sets, the powers of $2$ and the factorials are trivially bounded, and so are unions of intervals
$[a_k,b_k]$ with $a_{k+1}>U(b_k)$, which can have upper density $1$. So it is the gaps of $S$, not
its density, that matter here.

Most natural infinite sets are not trivially bounded. If $S$ has density $\delta>0$, then
$\#(S\cap(2B/3,B])=\delta B/3+o(B)$, so $S$ has only finitely many barriers. The odd numbers have
the single barrier $2$, the even numbers the barriers $1$ and $3$, and the squares the barriers
$2,3,6,7,8,14,15,24$. For the primes, Nagura's theorem \cite{Nag52} gives a prime in $(x,6x/5]$
for every $x\ge25$, so for $B\ge38$ there is a prime in $(2B/3,4B/5]\subseteq(2B/3,B]$, and a
check of $B\le37$ leaves only the barrier $B=1$.

The bound in (d) is sharp up to the constant. The chain of $1$, the set
$\{U^r(1):r\ge0\}=\{1,2,3,5,8,12,18,27,\dots\}$, has $s_{i+1}=U(s_i)$ for all $i$, so it is not trivially bounded, and it has $\log_{3/2}N+O(1)$
elements below $N$. Under its map every orbit diverges: a number outside the set goes down, so its
orbit decreases until it meets the set (at the latest at $1$), and from then on it goes up forever.

\subsection{Chains and runs}\label{sec:runs}

If $S$ is not trivially bounded, every large $B$ is jumped over: some element of $S$ in $(2B/3,B]$
goes up to a number above $B$. To describe these sets we use the structure of the step up alone.
By \cref{lem:steps}(a) and (e), the map $U$ is strictly increasing and its image is the set of
$y\ge2$ with $y\not\equiv1\pmod 3$. Hence $\N$ is the disjoint union of the \emph{chains}
\[
\Ch_c=\{c,\,U(c),\,U^2(c),\,\dots\},\qquad c\equiv1\pmod 3,
\]
which are the orbits of $T_\N$ and the rays of $G_\N$. For example,
\[
\Ch_1=\{1,2,3,5,8,12,18,27,\dots\},\quad \Ch_4=\{4,6,9,14,21,32,\dots\},\quad
\Ch_7=\{7,11,17,26,39,\dots\}.
\]
Every $x$ is $U^d(c)$ for a unique \emph{root} $c\equiv1\pmod3$ and a unique \emph{depth} $d=d(x)$:
the depth is the number of times that $U^{-1}$ can be applied to $x$. For instance $21=U^4(4)$ has
root $4$ and depth $4$. For any $x$, the set $\{x,U(x),U^2(x),\dots\}$ is the part of the chain of
$x$ from $x$ on; we call it a \emph{tail} of the chain.

An orbit of $T_S$ that arrives at an element $x$ of $S$ goes up to $U(x)$, and it goes up again
exactly if $U(x)\in S$. So the steps up of an orbit come in runs, and a run is a segment of a
chain that lies in $S$. For $x\in S$ let $\run_S(x)\in\{1,2,\dots,\infty\}$ be the largest $\run$
such that $x,U(x),\dots,U^{\run-1}(x)$ all lie in $S$; we call this number the \emph{run} of $x$ in
$S$. An orbit that arrives at $x\in S$ makes exactly $\run_S(x)$ steps up before its next step
down. Let $\run(S)=\sup_{x\in S}\run_S(x)$, with $\run(\emptyset)=0$.

For example, let $S=\odd$, so that $T_S$ is the Collatz map. From $7$ the orbit goes up three
times, $7\to11\to17\to26$, and then down, because $7$, $11$, $17$ are odd and $26$ is even. So
$\run_S(7)=3$, $\run_S(11)=2$ and $\run_S(17)=1$. In general, for odd $x$ the run $\run_{\odd}(x)$
is the exponent of $2$ in $x+1$, since $U(x)+1=\frac32(x+1)$ for odd $x$. All these runs are
finite, but they are not bounded: $\run(\odd)=\infty$.

We call an orbit $x_0,x_1,x_2,\dots$ \emph{increasing} if it goes up at every step,
$x_0<x_1<x_2<\cdots$. An increasing orbit diverges, and it is the simplest kind of divergent
orbit. The first part of the next proposition says when it occurs: an orbit that always goes up
is an orbit of $U$, so all of its elements must lie in $S$.

\begin{proposition}[Runs]\label{prop:runs}
Let $S\subseteq\N$.
\begin{enumerate}[label=(\alph*)]
\item The orbit of $x$ under $T_S$ is increasing if and only if all the numbers
$x,U(x),U^2(x),\dots$ lie in $S$, that is, if and only if $x\in S$ and $\run_S(x)=\infty$.
Consequently, $T_S$ has an increasing orbit if and only if $S$ contains a tail of a chain.
\item If $\run(S)\le1$, that is, if $U(x)\notin S$ for every $x\in S$, then for every $x\ge4$
either $T_S(x)<x$ or $T_S^2(x)=\ceil{3x/4}<x$. Hence $T_S$ is bounded, and every orbit ends in
one of the cycles $(1)$, $(1,2)$, $(2,3)$, $(3,5)$; these are cycles of $T_S$ if and only if
$1\notin S$, $1\in S$, $2\in S$ and $3\in S$, respectively. In particular, $T_S$ has one or two
cycles.
\item For every $y\ge51$, the path from $y$ that repeats the steps up, up, down is a ray of $G$.
The set $S$ of the vertices at which it goes up has $\run(S)=2$, and the orbit of $y$ under $T_S$
diverges.
\end{enumerate}
\end{proposition}

Here is what the three parts say. For (a), take the chain
$\Ch_4=\{4,6,9,14,21,32,\dots\}$. The orbit of $4$ under $T_S$ is $4,6,9,14,21,32,\dots$ if and
only if $S$ contains all these numbers. If $S$ contains only the tail $9,14,21,32,\dots$ of this
chain, then the orbit of $9$ is increasing, and so is, from $9$ on, every orbit that reaches $9$.
If $S$ misses infinitely many elements of every chain, as the odd numbers do, then all runs are
finite and no orbit is increasing. Such a map can still have divergent orbits, which go down
infinitely often but go up more: (c) gives the simplest example, in which all runs have length
$2$. Part (b) says that runs of length $1$ are not enough for this. So the threshold lies between
runs of length one and two: a step up followed by a step down multiplies a number by about $3/4$,
while two steps up and one step down multiply it by about $9/8$. We return to this threshold in
\cref{sec:threshold}.

\begin{proof}
(a) If the orbit of $x$ is increasing, then each of its steps is a step up, so the orbit is
$x,U(x),U^2(x),\dots$, and each of these numbers lies in $S$, because $T_S$ goes up only at
elements of $S$. Conversely, if all $U^j(x)$ lie in $S$, then $T_S(U^j(x))=U^{j+1}(x)$ for all
$j\ge0$, so the orbit of $x$ is $x,U(x),U^2(x),\dots$, which is increasing. The last statement
follows, since the set $\{x,U(x),U^2(x),\dots\}$ is a tail of a chain.

(b) Let $x\ge4$. If $x\notin S$, then $T_S(x)=D(x)<x$. If $x\in S$, then $U(x)\notin S$, so
$T_S^2(x)=D(U(x))=\ceil{3x/4}$ by \cref{lem:steps}(b), and $\ceil{3x/4}\le x-1$ for $x\ge4$. So
from every $x\ge4$ the orbit reaches a smaller number in one or two steps, passing at most
through $U(x)$. Hence every orbit is bounded and reaches $[1,3]$, and every cycle has its least
element in $[1,3]$. If $1\notin S$, then $1$ is fixed; if $1\in S$, then $2=U(1)\notin S$ and the
orbit of $1$ is $(1,2)$. If $2\in S$, then $3\notin S$ and $(2,3)$ is a cycle; if $2\notin S$,
then $2$ goes to $1$. If $3\in S$, then $5\notin S$ and $(3,5)$ is a cycle; if $3\notin S$, then
$3$ goes to $2$. So every orbit ends in one of the four cycles, and the conditions are as stated.
Exactly one of $(1)$ and $(1,2)$ occurs. Of the other two cycles at most one occurs, since
$2\in S$ excludes $3=U(2)\in S$, and $(2,3)$ does not occur together with $(1,2)$, since $1\in S$
excludes $2=U(1)\in S$. So $T_S$ has one or two cycles.

(c) We have to show that the path never visits a number twice; then it is a ray, and by
\cref{prop:dict}(b) it is a divergent orbit of the map that goes up exactly at its up-vertices.
Let $y_0=y$ and $y_{k+1}=F(y_k)$, where $F=D\circ U\circ U$, so that the path is
$y_0$, $U(y_0)$, $U^2(y_0)$, $y_1$, $U(y_1)$, $U^2(y_1)$, \dots. By \cref{lem:steps}(a),
$\frac32z\le U(z)\le\frac32z+\frac12$, hence $\frac94z\le U^2(z)\le\frac94z+\frac54$ and
$\frac98z\le F(z)\le\frac98z+\frac98$. By induction,
\begin{equation}\label{eq:Fbounds}
(9/8)^jy_k\le y_{k+j}\le(9/8)^j(y_k+9)-9\qquad(j\ge0),
\end{equation}
and the three sequences $(y_k)$, $(U(y_k))$, $(U^2(y_k))$ are strictly increasing. Within a
block $y_k<U(y_k)<U^2(y_k)$. So a repetition can only be of the form $y_{k+j}=U(y_k)$,
$y_{k+j}=U^2(y_k)$ or $U(y_{k+j})=U^2(y_k)$ with $j\ge1$, and the third is the first, as $U$ is
injective. Let $z=y_k\ge y_0\ge51$. For $j\le3$, \eqref{eq:Fbounds} gives
$y_{k+j}\le(9/8)^3(z+9)-9<1.4239z+3.82<\frac32z\le U(z)$, and for $j\ge4$ it gives
$y_{k+j}\ge(9/8)^4z>1.6018z>\frac32z+\frac12\ge U(z)$. Similarly, for $j\le6$,
$y_{k+j}\le(9/8)^6(z+9)-9<2.0273z+9.25<\frac94z\le U^2(z)$, and for $j\ge7$,
$y_{k+j}\ge(9/8)^7z>2.2806z>\frac94z+\frac54\ge U^2(z)$. So the path is a ray. Its up-vertices
are the $y_k$ and $U(y_k)$. Since the ray is simple, $U^2(y_k)$ is not one of them, and $y_k$
is not $U$ of one of them. So the runs of $S$ are the pairs $\{y_k,U(y_k)\}$, and the orbit of $y$
under $T_S$ is the ray.
\end{proof}

A computation shows that the path of \cref{prop:runs}(c) is simple for every $y\ge37$, but
repeats a number for $y\le36$ unless $y\in\{29,33,35\}$.

The condition in (b) is easy to check for many sets. For the squares, $U(k^2)$ is either $3k^2/2$
or $(3k^2+1)/2$, and neither is a square, since $3k^2\neq2j^2$ and $3k^2+1=2j^2$ is impossible
modulo $3$. So $T_{\{\text{squares}\}}$ has the single cycle $(1,2)$, which every orbit reaches.
The same holds for $4\N-3=\{1,5,9,13,\dots\}$, since $U(4k+1)=6k+2$ is even, and also for
$\{3\}\cup(4\N-3)$, which has the run $3,5$ of length $2$: the argument of (b) only needs
$U(x)\notin S$ for the elements $x\ge4$ of $S$. The condition does not make $S$ sparse:

\begin{lemma}\label{lem:E}
Let $E=\{x\in\N: d(x)\text{ even}\}$. Every $S\subseteq\N$ with $\run(S)\le1$ satisfies
$\#(S\cap[1,N])\le\#(E\cap[1,N])$ for all $N$, and $E$ has density $3/5$. More generally, the
numbers of depth $d$ have density $\frac13(\frac23)^d$, and the numbers of depth at least $d$
have density $(\frac23)^d$.
\end{lemma}

\begin{proof}
On a chain $\Ch_c=\{c_0<c_1<\cdots\}$, with $c_i=U^i(c)$, a set $S$ with $\run(S)\le1$ contains no
two consecutive elements $c_i,c_{i+1}$. The elements of $\Ch_c$ in $[1,N]$ are $c_0,\dots,c_{q-1}$
for some $q$, of which $S$ contains at most $\lceil q/2\rceil$ and $E$ contains exactly
$\lceil q/2\rceil$. Summing over $c$ gives the inequality. The numbers of depth $d$ form the set
$U^d(R)$, where $R=\{c:c\equiv1\pmod3\}$. By \cref{lem:steps}(d), $U^d(z)\le N$ if
$z+1\le(2/3)^dN$, and $U^d(z)>N$ if $z>(2/3)^dN$. Since $U^d$ is increasing, the number of
elements of $U^d(R)$ in $[1,N]$ is $(1/3)(2/3)^dN+O(1)$, and the number of elements of depth at
least $d$ is at most $(2/3)^dN+1$. Hence the densities are as claimed, and $E$ has density
$\sum_{d\text{ even}}\frac13\bigl(\frac23\bigr)^d=\frac13\cdot\frac95=\frac35$.
\end{proof}

In words: $E$ takes every second element of every chain, starting with the root, and no set
without two steps up in a row can do better. Since $d(1)=0$, $d(2)=1$ and $d(3)=2$, we have
$1,3\in E$ and $2\notin E$, and \cref{prop:runs}(b) gives the cycles $(1,2)$ and $(3,5)$ for
$T_E$. For comparison, the odd numbers have the smaller density $1/2$, but their runs are
unbounded: $U^j(2^k-1)=3^j2^{k-j}-1$ is odd for $j<k$, so $\run_{\odd}(2^k-1)=k$. The same holds
for the even numbers, with $2^k$ in place of $2^k-1$.

\subsection{Increasing orbits from sparse periodic sets}

A chain has density zero and gives an increasing orbit. Remarkably, the set of all numbers of
large depth is periodic, so even periodic sets of small density can contain chain tails.

\begin{proposition}\label{prop:depth}
Let $j\ge1$, let $D_j=U^j(\N)$ be the set of numbers of depth at least $j$, and let
$A_j=\{U^j(r)\bmod 3^j:1\le r\le2^j\}$. Then $\#A_j=2^j$ and
\[
D_j=\{x\in\N:\ x\bmod 3^j\in A_j\},
\]
so $D_j$ is periodic modulo $3^j$ and has density $(2/3)^j$. Under $T_{D_j}$, every orbit that
meets $D_j$ increases from then on, and every other orbit decreases to the fixed point $1$.
\end{proposition}

For example, $D_1=\{x:x\not\equiv1\pmod3\}$ is the image of $U$. A number in $D_1$ goes up to a
number in the image of $U$, that is, again to a number in $D_1$, so it goes up forever; a number
$x\equiv1\pmod3$ goes down, and continues to go down until it meets $D_1$ or arrives at $1$. The
next case is $D_2=\{x:x\equiv0,3,5,8\pmod9\}$, of density $4/9$. The reason for the periodicity
is that adding $2^j$ to $x$ adds $3^j$ to $U^j(x)$.

\begin{proof}
By \cref{lem:steps}(f) and induction on $i$, $U^i(x+2^j)=U^i(x)+3^i2^{j-i}$ for $0\le i\le j$, so
$U^j(x+2^j)=U^j(x)+3^j$. Since $U^j$ is strictly increasing, $D_j$ is the union of the
progressions $U^j(r)+3^j\N_0$, $1\le r\le2^j$. The numbers $U^j(1)<\dots<U^j(2^j)$ lie in
$[1,3^j]$, as $U^j(2^j)=3^j$, so they are distinct modulo $3^j$ and $\#A_j=2^j$. If
$x\equiv U^j(r)\pmod{3^j}$ with $x\ge1$, then $x\ge U^j(r)$, since $x\le U^j(r)-3^j\le0$
otherwise; so $x\in D_j$. This proves the description of $D_j$, and its density is $2^j/3^j$.
If $x\in D_j$, then $U(x)\in D_{j+1}\subseteq D_j$, so an orbit that meets $D_j$ stays in $D_j$ and
increases. An orbit outside $D_j$ goes down until it meets $D_j$ or reaches $1$, which has depth
$0$ and is fixed.
\end{proof}

\subsection{One-way doors}

In the other direction, there is a simple mechanism that makes $T_S$ bounded even when $S$ is very
large. If the complement $W=\N\setminus S$ is closed under the step down, then an orbit that enters
$W$ goes down and stays in $W$ forever, so it falls to $1$. The only way for an orbit to escape
this trap is to avoid $W$ altogether, and then it goes up at every step. So $W$ works like a
one-way door: orbits can pass from $S$ into $W$, but not back.

\begin{proposition}\label{prop:door}
Let $W\subseteq\N$ be nonempty with $D(W)\subseteq W$, and let $S=\N\setminus W$.
\begin{enumerate}[label=(\alph*)]
\item $1\in W$, and $S$ is closed under $x\mapsto2x-1$ and $x\mapsto2x$. In binary, a nonempty set
$W$ is closed under $D$ if and only if the set $\{x-1:x\in W\}$ is closed under deleting the last
binary digit.
\item Every orbit of $T_S$ that meets $W$ stays in $W$ and reaches the fixed point $1$.
\item $T_S$ is bounded if and only if $S$ contains no tail of a chain, that is, if and only if
every chain has infinitely many elements in $W$. In that case every orbit of $T_S$ reaches $1$.
\end{enumerate}
\end{proposition}

\begin{proof}
(a) $D^s(x)=1$ for large $s$, by \cref{lem:steps}(b). The numbers $2x-1$ and $2x$ are the two
preimages of $x$ under $D$, so $D(W)\subseteq W$ is equivalent to the closure of $S$ under them.
Finally $D(x)-1=\floor{(x-1)/2}$. (b) On $W$ the map $T_S$ is $D$, which maps $W$ into itself, and
every orbit of $D$ reaches $1$. (c) An orbit that never meets $W$ lies in $S$, so it goes up at
every step: it is a tail of a chain contained in $S$. Conversely, a chain tail in $S$ is an
increasing orbit. Together with (b) this proves (c).
\end{proof}

\begin{example}\label{ex:Wm}
For $m\ge1$ let
\[
W_m=[1,2^m]\ \cup\ \bigcup_{n\ge m}\bigl(2^n,\,2^n+2^{n-m}\bigr],
\]
that is, $x\in W_m$ if and only if $x-1<2^m$ or the binary expansion of $x-1$ begins with a $1$
followed by $m$ zeros. Then $W_m$ is closed under $D$: if $2^n<x\le2^n+2^{n-m}$ with $n>m$, then
$2^{n-1}<D(x)\le2^{n-1}+2^{n-1-m}$, and $D$ maps $(2^m,2^m+1]$ and $[1,2^m]$ into $[1,2^m]$. Every
chain meets $W_m$ infinitely often: $x\in(2^n,2^n(1+2^{-m})]$ means that $\{\log_2x\}$ lies in
the interval $(0,\log_2(1+2^{-m})]$, and by \cref{lem:steps}(d) the numbers $\log_2U^j(c)$ differ
from $\log_2K_c+j\log_2\frac32$ by less than $3(2/3)^j$ for $j\ge2$, and the fractional parts of the latter are dense, since
$\log_2\frac32$ is irrational. So for $S_m=\N\setminus W_m$ the map $T_{S_m}$ is bounded, and every
orbit reaches $1$. A short count shows that $S_m$ has lower density $(2^m-1)/(2^m+1)$ and upper
density $1-2^{-m}$; for instance $S_1=\bigcup_{n\ge1}(3\cdot2^{n-1},2^{n+1}]$, the set of $x$ for
which $x-1$ begins with $11$ in binary, has lower density $1/3$ and upper density $1/2$.
\end{example}

In this example the door $W_m$ occupies the proportion $2^{-m}$ of every interval
$(2^n,2^{n+1}]$, and a chain cannot avoid it, because the numbers $U^j(c)$ grow like
$K_c(3/2)^j$ and their positions within the intervals $(2^n,2^{n+1}]$ are spread out evenly.
Letting the number $m$ of zeros grow slowly with $n$ makes the door thinner and thinner, and
gives a set $S$ of density one.

\begin{theorem}\label{thm:density1}
There is a set $W\subseteq\N$ of density zero that is closed under $D$ and contains infinitely
many elements of every chain. Consequently $S=\N\setminus W$ has density one, $T_S$ is bounded,
and every orbit of $T_S$ ends at the fixed point $1$.
\end{theorem}

\begin{proof}
For a non-decreasing function $h\colon\N_0\to\N_0$ with $h(n)\le n$ let
\[
W_h=\{1\}\cup\bigcup_{n\ge0}\bigl(2^n,\,2^n+2^{n-h(n)}\bigr].
\]
\emph{$W_h$ is closed under $D$.} We have $D(1)=D(2)=1$. Let $n\ge1$ and
$2^n<x\le2^n+2^{n-h(n)}$. Then $D(x)>2^{n-1}$ and
$D(x)\le2^{n-1}+\ceil{2^{n-1-h(n)}}\le2^{n-1}+2^{n-1-h(n-1)}$, since $h(n-1)\le h(n)$ and
$h(n-1)\le n-1$. So $D(x)\in W_h$.

\emph{If $h(n)\to\infty$, then $W_h$ has density zero.} The interval $(2^n,2^{n+1}]$ contains
$2^{n-h(n)}$ elements of $W_h$. For $2^n<N\le2^{n+1}$ and every $L\le n$, splitting the sum at
$n-L$ gives
\[
\#(W_h\cap[1,N])\le2+\sum_{i\le n-L}2^i+\sum_{n-L<i\le n}2^{i-h(n-L)}
\le2+2^{n-L+1}+2^{n+1-h(n-L)},
\]
so $\limsup_N\#(W_h\cap[1,N])/N\le2^{1-L}$ for every $L$.

\emph{Choice of $h$.} Let $\alpha=\log_2\frac32$, which is irrational, and let $c_1,c_2,\dots$ be
a sequence of numbers $\equiv1\pmod3$ in which every such number appears infinitely often. We
choose $0=L_0<L_1<L_2<\cdots$ and elements $x_i\in\Ch_{c_i}$ with
\[
2^{L_i}<x_i<2^{L_i}+2^{L_i-i},\qquad L_i>L_{i-1}+i,
\]
as follows. Let $\varepsilon=\log_2(1+2^{-i})$ and $\beta=\log_2K_{c_i}$. By
\cref{lem:steps}(d), $\log_2U^j(c_i)=\beta+j\alpha+\eta_j$ with
$\log_2\bigl(1-(2/3)^j\bigr)<\eta_j<0$. By Kronecker's theorem the fractional part of
$\beta+j\alpha$ lies in $(\varepsilon/3,2\varepsilon/3)$ for infinitely many $j$; choose such a
$j$ with $|\eta_j|<\varepsilon/3$ and $U^j(c_i)>2^{L_{i-1}+i+1}$, and put $x_i=U^j(c_i)$ and
$L_i=\floor{\log_2x_i}$. Then $\{\log_2x_i\}\in(0,\varepsilon)$, which is the displayed
condition. Now let $h(0)=0$ and $h(n)=i$ for $L_{i-1}<n\le L_i$. This function is non-decreasing,
tends to infinity, and satisfies $h(n)\le n$ because $L_{i-1}\ge i-1$. Since $h(L_i)=i$, every
$x_i$ lies in $W_h$. Every chain is $\Ch_{c_i}$ for infinitely many $i$, and the $x_i$ are
distinct, so every chain has infinitely many elements in $W=W_h$. The rest follows from
\cref{prop:door}.
\end{proof}

The function $h$ in the proof grows very slowly, and it could be made explicit with an effective
irrationality measure for $\log_23$; we do not need this.

\subsection{Density is not the measure}

The examples of this section show that the density of $S$ does not determine, even roughly,
whether $T_S$ is bounded:
\begin{itemize}
\item density zero is compatible with divergence: a single chain gives an increasing orbit, and
so do the periodic sets $D_j$ of density $(2/3)^j$ (\cref{prop:depth});
\item density one is compatible with convergence of every orbit to $1$ (\cref{thm:density1}),
and upper density one even with trivial boundedness;
\item in between, the set $E$ of density $3/5$ gives a bounded map with two small cycles
(\cref{lem:E}), while the odd and the even numbers, of density $1/2$, give the Collatz map.
\end{itemize}
What matters is how often an orbit actually visits $S$, and an orbit visits $S$ in a very biased
way: chains are the only routes upward, and a set can be dense but avoid them often enough, or
sparse but contain one. We take this up in \cref{sec:threshold}. \Cref{tab:classes} collects the
classes found so far; they are not exclusive.

\begin{table}[ht]
\centering
\caption{The simplest classes of sets $S$. The classes overlap; for instance, sets whose complement
is closed under $D$ can be trivially bounded.}\label{tab:classes}
\begin{tabular}{@{}>{\raggedright\arraybackslash}p{0.25\textwidth}>{\raggedright\arraybackslash}p{0.25\textwidth}>{\raggedright\arraybackslash}p{0.44\textwidth}@{}}
\toprule
class & condition & the map $T_S$\\
\midrule
finite & & bounded; $[1,U(\max S)]$ is invariant\\
trivially bounded & infinitely many barriers & bounded, and so is $T_{S'}$ for every $S'\subseteq S$\\
\multicolumn{3}{@{}l}{not trivially bounded:}\\
\quad no two steps up in a row & $\run(S)\le1$ & bounded, with one or two of the cycles $(1)$, $(1,2)$, $(2,3)$, $(3,5)$; the density can be $3/5$\\
\quad one-way doors & $D(\N\setminus S)\subseteq\N\setminus S$ & bounded if and only if $S$ contains no chain tail, and then every orbit reaches $1$; the density can be $1$\\
\quad an infinite run & $\run_S(x)=\infty$ for some $x$ & unbounded, with an increasing orbit; the set can be periodic with density $(2/3)^j$\\
\quad all other sets & all runs finite, $\run(S)\ge2$ & bounded for $\{3\}\cup(4\N-3)$, unbounded in \cref{prop:runs}(c); open for the odd and the even numbers and the primes\\
\bottomrule
\end{tabular}
\end{table}

\subsection{Typical sets}

Two natural notions of a typical set give opposite answers. The second is that of a random set.
The first is topological, and we recall the terms. A set $S\subseteq\N$ is the same thing as a
point of the product $\{0,1\}^\N$, and we use the \emph{product topology} on the collection of all
sets: a family $\mathcal U$ of sets is \emph{open} if for every $S\in\mathcal U$ there is an $n$
such that $\mathcal U$ contains every set $S'$ with $S'\cap[1,n]=S\cap[1,n]$. In other words,
membership in an open family is always witnessed by finitely many elements and non-elements of the
set. A family is \emph{dense} if it meets every nonempty open family, that is, if every
prescription of finitely many elements and non-elements is satisfied by one of its members. A
\emph{$G_\delta$} is an intersection of countably many open families. By the Baire category
theorem, an intersection of countably many dense open families is dense; these intersections are
the \emph{dense $G_\delta$} families. A family is \emph{comeager} if it contains a dense
$G_\delta$. Comeager families are dense, in particular nonempty, and an intersection of countably
many comeager families is comeager; see \cite{Oxt80}. So the comeager families play the role of
``almost all sets'' in the topological sense, as the families of probability one do for random
sets, and a property is called \emph{generic} if the sets that have it form a comeager family.

\begin{proposition}[Typical sets]\label{prop:typical}
\leavevmode
\begin{enumerate}[label=(\alph*)]
\item In the product topology on the collection of all sets $S\subseteq\N$, the trivially bounded
sets form a dense $G_\delta$, and so do the sets $S$ for which $T_S$ has infinitely many cycles. So
for a comeager family of sets $S$ the map $T_S$ is trivially bounded and has infinitely many
cycles.
\item If every $x$ belongs to $S$ independently with probability $p\in(0,1)$, then almost surely
$S$ is not trivially bounded, all its runs are finite, and $\run(S)=\infty$. If $p<\log_32$, then
$T_S$ is almost surely bounded.
\end{enumerate}
\end{proposition}

In (b) the logarithm of an orbit behaves, until the orbit returns to a number it has visited, like
a random walk with steps $\log(3/2)$ and $-\log2$, which has negative drift for $p<\log_32$. We
expect that $T_S$ almost surely has divergent orbits if $p>\log_32$ (\cref{q:random}). The proof
of (a) uses \cref{thm:G}(b) from the next section.

\begin{proof}
(a) The set $\{S: S\cap(2B/3,B]=\emptyset\}$ is closed and open, and for every $n$ the union of
these sets over $B\ge n$ is open and dense: a condition on finitely many elements can be met by a
set that has no element in $(2B/3,B]$ for some large $B$. So the trivially bounded sets form a
dense $G_\delta$. The set of $S$ for which $T_S$ has at least $k$ cycles is open, since a cycle
depends only on $S$ at its elements, and dense, since by \cref{thm:G}(b) we can add $k$ cycles
whose least elements exceed the largest prescribed number and the largest element of the
previously added cycle. Intersecting over $k$ gives the second dense $G_\delta$.

(b) A given $B$ is a barrier with probability $(1-p)^{\#(2B/3,B]}\le(1-p)^{B/3}$, which is
summable, so by the Borel--Cantelli lemma $S$ has only finitely many barriers. A given chain tail
lies in $S$ with probability $0$, and there are countably many; and for every $\run$, disjoint
segments of length $\run$ of chains lie in $S$ independently with probability $p^\run$, so
infinitely many of them do. For the last claim fix $x_0$, and let $\xi_1,\xi_2,\dots$ be
independent with $P(\xi_t=1)=p=1-P(\xi_t=0)$. We may assume that the membership of the $t$-th new
number visited by the orbit of $x_0$ is decided by $\xi_t$, so that until the first repetition the
orbit makes a step up at time $t$ if and only if $\xi_t=1$. Suppose that the orbit never repeats
a number. Then $x_t\to\infty$, and since $U(z)\le\frac32z(1+\frac1{3z})$ and
$D(z)\le\frac z2(1+\frac1z)$,
\[
\log x_t\le\log x_0+\sum_{i=1}^t\bigl(\xi_i\log\tfrac32-(1-\xi_i)\log2\bigr)
+\sum_{i<t}\log\bigl(1+\tfrac1{x_i}\bigr).
\]
The last sum is $o(t)$, and by the strong law of large numbers the first sum is
$t\bigl(p\log3-\log2\bigr)+o(t)$ almost surely, which tends to $-\infty$ for $p<\log_32$. This
contradicts $x_t\to\infty$. So almost surely the orbit of $x_0$ repeats a number, for every
$x_0$.
\end{proof}

\section{Cycles across all maps}\label{sec:cycles}

By \cref{prop:dict}, the cycles of all maps $T_S$ together are the simple cycles of $G$, and a set
of simple cycles of $G$ occurs in a single map if and only if its members are pairwise disjoint.
So the question whether there are infinitely many cycles across all maps is the question whether
$G$ has infinitely many simple cycles.

\begin{theorem}\label{thm:G}
\leavevmode
\begin{enumerate}[label=(\alph*)]
\item $G$ is strongly connected. More precisely, for all $x\ge1$ and $y\ge2$ there are $r,s\ge1$
with $D^s(U^r(x))=y$.
\item Every $n\in\N$ is the least element of a cycle of some map $T_S$, and for $n\ge2$ the
cycle can be chosen with a single run of steps up. In particular, there are infinitely many
cycles across all maps.
\item Let a cycle have length $\ell$, $a$ steps up and least element $m$. Then
$1<2^\ell/3^a\le(1+1/m)^\ell$. Hence there are only finitely many cycles of each length.
\item If $p_1,\dots,p_k\ge2$ are the numbers at which the runs of steps up of a cycle other than
$(1)$ start, then
\[
\frac{3^a}{2^\ell}>\prod_{j=1}^k\frac{p_j-1}{p_j+1}.
\]
In particular, every cycle with a single run of steps up has $3^a<2^\ell<3^{a+1}$, that is,
$a=\floor{\ell\log_32}$.
\end{enumerate}
\end{theorem}

The idea behind (a) and (b) is that a run of steps up multiplies by almost exactly $(3/2)^r$ and a
run of steps down divides by almost exactly $2^s$, and since $\log 3/\log 2$ is irrational,
$(3/2)^r/2^s$ can be made as close to any given ratio as we like. Going up from $n$ and coming back
down to $n$ closes a cycle. For example, from $n=4$ the steps up give $4,6,9,14,21,\dots$, and we
look for an element of this chain that the steps down bring back to $4$, that is, an element
$u$ with $\ceil{u/2^s}=4$ for some $s$. The number $6$ fails ($6\to3$), and so does $9$
($9\to5\to3$), but $14\to7\to4$ works. This gives the cycle $(4,6,9,14,7)$ of length $\ell=5$ with
$a=3$ steps up, which is a cycle of $T_S$ exactly for the sets $S$ that contain $4,6,9$ and do
not contain $7,14$.

Part (c) is the identity that every cycle of every map must satisfy,
the analogue of the well-known product formula for Collatz cycles: the steps multiply by $3/2$
and $1/2$ exactly at even numbers and by slightly more at odd numbers, and around a cycle the
total factor is $1$. For the cycle $(4,6,9,14,7)$ the odd elements are $9$, which goes up, and
$7$, which goes down, and indeed
$2^5/3^3=\frac{32}{27}=\bigl(1+\frac1{27}\bigr)\bigl(1+\frac17\bigr)$. Part (d) is a relation of
Steiner's type \cite{Ste78} that holds for all maps at once.

\begin{lemma}\label{lem:reach}
For all $x\ge1$ and $y\ge2$ there are infinitely many pairs $r,s\ge1$ with $D^s(U^r(x))=y$.
\end{lemma}

\begin{proof}
By \cref{lem:steps}(b), $D^s(U^r(x))=y$ if and only if $y-1<U^r(x)/2^s\le y$. Let
$J=\bigl(\log_2(y-\tfrac12),\log_2y\bigr)$ and $\lambda_r=\log_2K_x+r\log_2\frac32$, with $K_x$ as
in \cref{lem:steps}(d). If $\lambda_r-s\in J$ and $s\ge1$, then $K_x(3/2)^r/2^s\in(y-\frac12,y)$,
so $U^r(x)/2^s\in(y-\frac12-2^{-s},y)\subseteq(y-1,y)$. The interval $J$ has length less than $1$
and contains no integer, since no power of $2$ lies in $(y-\frac12,y)$; so $J\subseteq(j,j+1)$ for
$j=\floor{\log_2(y-\frac12)}$. Since $\log_2\frac32$ is irrational, the fractional part of
$\lambda_r$ lies in $J-j$ for infinitely many $r$, by Kronecker's theorem, and for each of them
$s=\floor{\lambda_r}-j$ satisfies $\lambda_r-s\in J$, and $s\ge1$ when $r$ is large.
\end{proof}

\begin{proof}[Proof of \cref{thm:G}]
(a) Every $x$ reaches $1$ by steps down, since $D^s(x)=\ceil{x/2^s}$, and $1$ reaches every
$y\ge2$ by \cref{lem:reach}.

(b) For $n=1$ take the fixed point $(1)$. Let $n\ge2$ and choose $r,s\ge1$ with
$D^s(U^r(n))=n$. Put $u_i=U^i(n)$ for $0\le i\le r$ and $v_j=D^j(u_r)$ for $0\le j\le s$. Then
$n=u_0<u_1<\dots<u_r=v_0>v_1>\dots>v_s=n$, and $v_j>n$ for $j<s$, since
$D(v_{s-1})=n$ gives $v_{s-1}\ge2n-1>n$. So the walk goes up from $n$ to $u_r$ and comes down to
$n$ again; the only way in which it can fail to be a simple cycle is that on the way down it
meets a number that it visited on the way up, and then we take a shortcut. If no $v_j$ with $1\le j<s$ is one of $u_1,\dots,u_r$, the
closed walk $u_0,\dots,u_r,v_1,\dots,v_{s-1}$ is a simple cycle. Otherwise let $j^*<s$ be the
largest $j$ for which it is, say $v_{j^*}=u_{i^*}$ with $i^*\ge1$; then
$u_0,\dots,u_{i^*},v_{j^*+1},\dots,v_{s-1}$ is a simple cycle. In both cases the cycle has least
element $n$ and a single run of steps up, and by \cref{prop:dict}(a) it is a cycle of a map
$T_S$. Since every cycle is finite and every $n$ lies on one, there are infinitely many cycles.

(c) Let $x_0,\dots,x_{\ell-1}$ be the cycle, with indices modulo $\ell$. By \cref{lem:steps}(a),
$x_{i+1}=(c_ix_i+e_i)/2$ with $c_i\in\{1,3\}$ and $e_i=x_i\bmod2$. Multiplying the ratios
$x_{i+1}/x_i$ around the cycle gives
\begin{equation}\label{eq:product}
\frac{2^\ell}{3^a}=\prod_{i=0}^{\ell-1}\Bigl(1+\frac{e_i}{c_ix_i}\Bigr).
\end{equation}
Not all $x_i$ are even, since both steps divide an even number by $2$ exactly once, so that
otherwise the exponent of $2$ in $x_i$ would decrease by $1$ at every step around the cycle. So the product is
larger than $1$, and every factor is at most $1+1/m$. For fixed $\ell$ there are finitely many
$a<\ell\log_32$, so $2^\ell/3^a$ stays above some constant larger than $1$; then the inequality
$2^\ell/3^a\le(1+1/m)^\ell$ bounds $m$, and a cycle is determined by $m$ and its $\ell$ steps.

(d) Write the cycle as runs of steps up $p_j\to q_j=U^{r_j}(p_j)$ and runs of steps down
$q_j\to p_{j+1}=D^{s_j}(q_j)$, $j=1,\dots,k$, with $p_{k+1}=p_1$, $\sum r_j=a$ and
$\sum(r_j+s_j)=\ell$. By \cref{lem:steps}(d), $q_j+1\le(3/2)^{r_j}(p_j+1)$, and by
\cref{lem:steps}(b), $p_{j+1}<q_j/2^{s_j}+1$. Hence
$p_{j+1}-1<(3/2)^{r_j}(p_j+1)/2^{s_j}$, and multiplying over $j$ gives
$\prod_j(p_j-1)<(3^a/2^\ell)\prod_j(p_j+1)$. If $k=1$ and $p_1\ge2$, this gives
$3^a/2^\ell>(p_1-1)/(p_1+1)\ge1/3$. If $k=1$ and $p_1=1$, the cycle is $(1,2)$, since the only
edges into $1$ come from $1$ and $2$; and $4<9$. Together with (c), $3^a<2^\ell<3^{a+1}$.
\end{proof}

Part (b) contrasts with Steiner's theorem that the only Collatz cycle with a single run of odd
steps is $(1,2)$ \cite{Ste78}, extended by Simons and de Weger to cycles with up to $75$ runs
\cite{SdW05}. The cycles of (b) are long, since $(3/2)^r/2^s$ has to be within about $1/n$ of $1$;
for $n=32$ the construction gives a cycle of length $46$ with largest element $4{,}102{,}266$.
Cycles with several runs are much shorter: for $88\%$ of the numbers $n$ between $9\cdot10^4$
and $10^5$ there is a cycle through $n$ of length at most $65$ inside $[1,3n]$.

Since a single map can have any set of disjoint cycles, the cycles of single maps carry no
further restriction:

\begin{proposition}[Prescribed cycles]\label{prop:prescribe}
Let $\mathcal C$ be a nonempty set, finite or infinite, of pairwise vertex-disjoint simple cycles
of $G$. Then there is a set $S$ such that the cycles of $T_S$ are exactly the members of
$\mathcal C$ and every orbit of $T_S$ ends in one of them.
\end{proposition}

\begin{proof}
The idea is to let every number outside the given cycles go down. Then every orbit decreases
until it meets one of the cycles or arrives at $1$, and we only have to take care of $1$.
Let $V$ be the union of the vertex sets of the members of $\mathcal C$, and $V^\uparrow$ the set of
vertices at which they go up. If $1\in V$, let $S=V^\uparrow$. Otherwise, let $P$ be a shortest
path from $2$ to $V$, which exists by \cref{thm:G}(a). It does not pass through $1$, since from
$1$ one can only go to $1$ or $2$, and only its last vertex lies in $V$. Let $S$ consist of
$V^\uparrow$, the number $1$, and the vertices at which $P$ goes up. Then $T_S$ follows the cycles
of $\mathcal C$ on $V$, sends $1$ to $2$ and then along $P$ into $V$, and sends every other number
$x$ down to $D(x)<x$. So every orbit decreases until it meets $V$, $P$ or $1$, and then ends in a
member of $\mathcal C$.
\end{proof}

So for every $k\in\{1,2,\dots,\infty\}$ there are bounded maps with exactly $k$ cycles, and $T_\N$
has none. The classification of single maps by their cycles is therefore the classification of
sets of pairwise disjoint simple cycles of $G$, and what remains is the structure of $G$ and of
natural classes of sets. \Cref{tab:census} lists the cycles of $G$ of length at most $30$.

\begin{table}[ht]
\centering
\caption{All simple cycles of $G$, that is, all cycles of all maps $T_S$, of length $\ell\le30$.
There are none of the lengths $3,4,6,7,9,11,12,14,17$--$20,22,23,25,28,30$. Every cycle found has
$a=\floor{\ell\log_32}$ steps up.}\label{tab:census}
\begin{tabular}{@{}rrrrrr@{}}
\toprule
$\ell$ & $a$ & $2^\ell/3^a$ & cycles & with one run of steps up & largest element\\
\midrule
1 & 0 & 2.000 & 1 & 0 & 1\\
2 & 1 & 1.333 & 3 & 3 & 5\\
5 & 3 & 1.185 & 12 & 6 & 26\\
8 & 5 & 1.053 & 140 & 21 & 242\\
10 & 6 & 1.405 & 18 & 2 & 41\\
13 & 8 & 1.249 & 198 & 0 & 89\\
15 & 9 & 1.665 & 12 & 0 & 41\\
16 & 10 & 1.110 & 2{,}056 & 10 & 912\\
21 & 13 & 1.315 & 2{,}283 & 2 & 1{,}065\\
24 & 15 & 1.169 & 144{,}932 & 8 & 5{,}165\\
26 & 16 & 1.559 & 4{,}975 & 1 & 1{,}598\\
27 & 17 & 1.039 & 8{,}066{,}988 & 28 & 47{,}201\\
29 & 18 & 1.386 & 302{,}979 & 2 & 6{,}245\\
\bottomrule
\end{tabular}
\end{table}

The three cycles of length $2$ are $(1,2)$, $(2,3)$ and $(3,5)$, which appeared in
\cref{prop:runs}(b). The table suggests that cycles exist for the lengths $\ell$ for which
$2^\ell/3^{\floor{\ell\log_32}}$ is small, but not in a monotone way (\cref{q:lengths}).

\begin{remark}[Cycles and divergent orbits together]\label{rem:types}
Call two divergent orbits of $T_S$ equivalent if they eventually coincide, and let $k$ and $l$ be
the numbers of cycles and of classes of divergent orbits of $T_S$; always $k+l\ge1$, since the
orbit of $1$ is periodic or diverges. Every pair $(k,l)$ with $k<\infty$ and $k+l\ge1$ occurs: take
$k$ disjoint cycles below some $K$, one of them $(1)$, and $l$ chains with roots above $K$ (or
$\Ch_1$ together with $l-1$ such chains if $k=0$), and let all other numbers go down. In this
construction the chains, the cycles and the number $1$ are disjoint, every number that lies on
none of them goes down, and so its orbit decreases until it meets one of them; distinct chains
are disjoint and are never left, so they give distinct classes of divergent orbits.
\Cref{prop:prescribe} gives $k=\infty$, $l=0$. We do not know whether a map can have infinitely
many cycles and a divergent orbit (\cref{q:types}).
\end{remark}

\section{Near the threshold}\label{sec:threshold}

Which sets are, intuitively, the most expanding sets whose maps can still be proved to be
bounded, and which are the least expanding sets whose maps can be proved to be unbounded? Since
the density of $S$ is not a measure of expansion (\cref{sec:classification}), we first need a
better yardstick. The natural one is how often an orbit actually goes up. In this section we show
that the critical proportion of steps up is $\log_32$, sharp from both sides, and that for sets
that are periodic modulo a power of $2$ the proportion of steps up of a typical orbit can be
computed from a finite Markov chain. This leads to a simple test, uniform descent, that proves
boundedness for many periodic sets, and it places Mahler's $Z$-number problem inside our family.

\subsection{The critical proportion of steps up}

Along an orbit, a step up multiplies by $3/2$ and a step down by $1/2$, up to a relative error of
order $1/x$. So after $t$ steps, of which $a$ go up, an orbit has been multiplied by about
$3^a/2^t$, and it grows or shrinks according to whether $a/t$ is above or below
$\log_32=0.6309\ldots$.

\begin{proposition}\label{prop:critical}
Let $x_0,x_1,\dots$ be an orbit of a map $T_S$, and let $a(t)$ be the number of steps up among its
first $t$ steps.
\begin{enumerate}[label=(\alph*)]
\item $x_0\,3^{a(t)}/2^t\le x_t\le x_0\,(3^{a(t)}/2^t)\prod_{i<t}(1+1/x_i)$.
\item If the orbit diverges, then $\liminf_{t\to\infty}a(t)/t\ge\log_32$.
\item If $\liminf_{t\to\infty}a(t)/t>\log_32$, then the orbit diverges.
\item Every cycle has a proportion $a/\ell<\log_32$ of steps up, and there are cycles with
proportions arbitrarily close to $\log_32$.
\end{enumerate}
\end{proposition}

\begin{proof}
(a) For every $z$, $\frac32z\le U(z)\le\frac32z(1+\frac1z)$ and
$\frac12z\le D(z)\le\frac12z(1+\frac1z)$; multiply over the first $t$ steps. (b) If the orbit
diverges, then $x_i\to\infty$, so $\sum_{i<t}\log(1+1/x_i)=o(t)$, and by (a)
$\log x_t\le\log x_0+a(t)\log3-t\log2+o(t)$. If $a(t)\le(\log_32-\varepsilon)t$ for infinitely many
$t$, then $\log x_t\to-\infty$ along these $t$, which is impossible. (c) By (a),
$x_t\ge x_0\exp\bigl(t(\tfrac{a(t)}t\log3-\log2)\bigr)\to\infty$. (d) The first claim is
\cref{thm:G}(c). By \cref{thm:G}(b), for every $n\ge2$ there is a cycle with least element $n$ and
one run of steps up, and by \cref{thm:G}(d) it has $a=\floor{\ell\log_32}$. Since there are only
finitely many cycles of each length, $\ell\to\infty$ as $n\to\infty$, and
$\floor{\ell\log_32}/\ell\to\log_32$.
\end{proof}

So no set $S$ can produce a divergent orbit that goes up less often than a proportion $\log_32$
of the time, in the long run. The next theorem shows that this bound is sharp: there are divergent
orbits that go up only very slightly more often. For the Collatz map, (b) is the familiar
observation that a divergent orbit would have to consist of odd numbers at least a proportion
$\log_32$ of the time.

\begin{theorem}[Barely divergent orbits]\label{thm:barely}
Let $r,s\ge1$ with $3^r>2^{r+s}$, that is, $r/(r+s)>\log_32$. There is $y_0$ such that for every
$y\ge y_0$ the path from $y$ in $G$ that repeats $r$ steps up followed by $s$ steps down is a ray.
If $S$ is the set of its up-vertices, then the orbit of $y$ under $T_S$ is this ray: it diverges,
and its proportion of steps up is $r/(r+s)$. In particular, there are divergent orbits whose
proportion of steps up is arbitrarily close to $\log_32$.
\end{theorem}

\Cref{prop:runs}(c) is the case $r=2$, $s=1$. The best choices of $(r,s)$ come from the
continued fraction of $\log_32=[0;1,1,1,2,2,3,1,5,2,23,\dots]$, whose convergents above $\log_32$
are $2/3$, $12/19$, $53/84$, $665/1054,\dots$; see \cref{tab:barely}.

\begin{table}[ht]
\centering
\caption{Barely divergent rays (\cref{thm:barely}): the pattern $U^rD^s$ repeated forever. The
growth per block is $\rho=3^r/2^{r+s}$, and $y_0$ is the threshold given by the proof.}\label{tab:barely}
\begin{tabular}{@{}rrlll@{}}
\toprule
$r$ & $s$ & $r/(r+s)$ & $\rho$ & $y_0$ from the proof\\
\midrule
$2$ & $1$ & $0.66667$ & $1.12500$ & $1{,}993$ (a computation gives $37$)\\
$12$ & $7$ & $0.631579$ & $1.013643$ & $3.3\cdot10^7$\\
$53$ & $31$ & $0.6309524$ & $1.002090$ & $6.3\cdot10^{11}$\\
\midrule
\multicolumn{2}{@{}l}{$\log_32$} & $0.6309298$ & $1$ &\\
\bottomrule
\end{tabular}
\end{table}

The path of the theorem grows by the factor $\rho=3^r/2^{r+s}>1$ in every block of $r+s$ steps,
so it tends to infinity. We have to prove that it never visits a number twice, because
only then is it the orbit of a single map (\cref{prop:dict}). The idea is this: Any two numbers on
the path differ by a factor $3^X/2^Y$, up to a relative error of order $1/y$ that comes from the
rounding at odd numbers. A power of $3$ is never a power of $2$, so $3^X/2^Y\ne1$; and only
finitely many exponents $X,Y$ have to be considered, because after many blocks the path has grown
too much to return. So all these factors stay away from $1$ by a fixed amount, and for large $y$
the rounding errors are too small to make two numbers of the path equal.

\begin{proof}
Let $L=r+s$, $\rho=3^r/2^L>1$, and let $z_0=y,z_1,z_2,\dots$ be the path, so that the $t$-th step
goes up if and only if $t\bmod L<r$. Let $a(t)$ be the number of steps up among the first $t$
steps. Since the steps up come first in every block, $3^{a(p)}/2^p\ge1$ for $0\le p\le L$, and so
$3^{a(t)}/2^t\ge\rho^{\floor{t/L}}$ for all $t$. By the bounds in the proof of
\cref{prop:critical}(a), $z_t\ge y\rho^{\floor{t/L}}$, and therefore
\[
z_t=y\,\frac{3^{a(t)}}{2^t}\,(1+\vartheta_t),\qquad 0\le\vartheta_t\le
\exp\Bigl(\sum_{i\ge0}\frac1{z_i}\Bigr)-1\le e^{C_1/y}-1,\qquad C_1=\frac{L\rho}{\rho-1}.
\]
Suppose that $z_t=z_{t'}$ with $t<t'$, and write $t'-t=qL+e$ with $0\le e<L$. Then
$a(t')-a(t)=qr+A$, where $0\le A\le e$ is the number of steps up among $e$ consecutive positions
of the periodic pattern, and
\[
Q:=\frac{3^{a(t')-a(t)}}{2^{t'-t}}=\rho^q\,\frac{3^A}{2^e}=\frac{1+\vartheta_t}{1+\vartheta_{t'}}
\in\bigl[e^{-C_1/y},e^{C_1/y}\bigr].
\]
Since $3^A/2^e\ge2^{-L}$, we have $Q>2$ as soon as $q>q_0=(L+1)/\log_2\rho$. For $q\le q_0$ there
are only finitely many possible values of $Q$, all of the form $3^X/2^Y$ with $Y=qL+e\ge1$, so all
different from $1$. Let $c>0$ be the smallest of the numbers $|\log Q|$ over these finitely many
values. Then for $y>y_0=C_1/c$ (and $y\ge C_1/\log2$, so that $e^{C_1/y}\le2$) no repetition is
possible, and the path is a ray. Its up-vertices are visited once each, so by
\cref{prop:dict}(b) the orbit of $y$ under $T_S$ follows the ray.
\end{proof}

\subsection{Periodic sets and fresh binary digits}\label{sec:fresh}

For a general set $S$ the sequence of steps of an orbit can be anything. For a set that is
periodic modulo a power of $2$, it is governed by a finite graph.

\begin{definition}\label{def:residue}
Let $k\ge1$, $M=2^k$, and let $S$ be periodic modulo $M$; we also regard $S$ as a set of residues
modulo $M$. For a residue $r$ let $\hat r\in[1,M]$ be its representative, and let
$g(r)=U(\hat r)\bmod M$ if $r\in S$ and $g(r)=D(\hat r)\bmod M$ otherwise. The \emph{residue graph}
$\Res_k(S)$ is the directed graph on $\Z/M$ with the two edges $r\to g(r)$ and $r\to g(r)+M/2$ at
every $r$. We give the vertex $r$ the factor $f(r)=3/2$ if $r\in S$ and $f(r)=1/2$ otherwise. The
\emph{residue chain} is the random walk on $\Res_k(S)$ that takes each of the two edges with
probability $1/2$.
\end{definition}

The two edges at $r$ are always different, since $M/2\not\equiv0\pmod M$. They lead to the two
possible residues of $T_S(x)$ for the numbers $x\equiv r\pmod M$. Indeed, by \cref{lem:steps}(f),
adding $M$ to $x$ adds $\frac32M$ to a step up and $\frac12M$ to a step down, and both are
congruent to $M/2$ modulo $M$; so the images of $\hat r,\hat r+M,\hat r+2M,\dots$ have the
residues $g(r)$ and $g(r)+M/2$ alternately (\cref{lem:fresh}(a)). In binary terms, a step ends
with a division by $2$, so the last $k$ binary digits of $x$ determine only the last $k-1$ digits
of $T_S(x)$.

\begin{example}\label{ex:residue4}
For the Collatz set $S=\odd$ and $k=2$, the residue graph on $\{0,1,2,3\}$ has the edges
\[
0\to0,2;\qquad 1\to0,2;\qquad 2\to1,3;\qquad 3\to1,3,
\]
with steps up at $1$ and $3$; for instance $U(3)=5\equiv1$, so $3\to1$ and $3\to1+2=3$. The loop
$3\to3$ has one step up in one step, and $3>2$; it reflects the fact that the numbers
$x\equiv-1\pmod{2^j}$ go up $j$ times in a row. The cycle $1\to2\to1$ is the residue pattern of
the cycle $(1,2)$, with $3<4$, and the cycle $2\to3\to1\to2$ has $9>8$. Every vertex has two
incoming edges, so the uniform distribution is stationary and $\pi=1/2$. For $S=4\N-3$, the only
change is that $3$ goes down, $D(3)=2$, so $3\to0,2$; now every cycle has at most half of its
vertices in $S$, and the stationary distribution $(\frac13,\frac16,\frac13,\frac16)$ gives
$\pi=1/6$.
\end{example}

The next lemma says that the residues of an orbit move along this graph, and that the choice
between the two edges is made by a new binary digit of the starting number at every step.

\begin{lemma}[Fresh binary digits]\label{lem:fresh}
Let $S$ be periodic modulo $M=2^k$.
\begin{enumerate}[label=(\alph*)]
\item If $x\equiv r\pmod M$, then $T_S(x)\bmod M$ is one of the two successors of $r$ in
$\Res_k(S)$, and $T_S(x)=f(r)\,x+O(1)$; precisely, $x=\hat r+Mu$ gives
$T_S(x)=T_S(\hat r)+2f(r)\cdot\frac M2u\equiv g(r)+\frac M2u\pmod M$.
\item Let $t\ge0$ and $r_0\in\Z/M$. The sequence of residues $(T_S^i(x)\bmod M)_{0\le i\le t}$
depends only on $x\bmod2^{k+t}$, and as $x$ runs through the $2^t$ classes modulo $2^{k+t}$ that
are $\equiv r_0\pmod M$, this sequence runs through all $2^t$ paths of length $t$ in $\Res_k(S)$
that start at $r_0$, each exactly once.
\item Consequently, for every path $P$ of length $t$ in $\Res_k(S)$ the set of $x\in\N$ whose
residue sequence begins with $P$ is a single residue class modulo $2^{k+t}$, of density
$2^{-k-t}$. In other words, the residues of the orbit of a number $x$ chosen uniformly at random
from $[1,2^{k+t}]$ form the first $t$ steps of the residue chain started from the uniform
distribution.
\end{enumerate}
\end{lemma}

For the Collatz map ($S=\odd$, $k=1$) this is Terras's observation that the sequence of parities of
the first $t$ terms of an orbit is a bijective function of $x\bmod2^t$ \cite{Ter76,Lag85}.

Here is what the lemma says in the graph of \cref{ex:residue4}. A number $x\equiv3\pmod4$ goes up
to $(3x+1)/2$, and its residue moves along one of the edges $3\to1$ and $3\to3$. Which one is
decided by the next binary digit of $x$: the numbers $x\equiv3\pmod8$, such as $3\to5$ and
$11\to17$, arrive at the residue $1$, and the numbers $x\equiv7\pmod8$, such as $7\to11$ and
$15\to23$, arrive at the residue $3$. So each edge is used by exactly half of the numbers, and
the digit that decides was not used before. At the next step the fourth binary digit of $x$
decides, and so on. For a number taken at random, the residues therefore perform a random walk on
the residue graph, with a fair coin tossed at every step, for as many steps as the number has
binary digits. This is why the properties of the map $T_S$ for typical numbers can be read off a
finite Markov chain.

\begin{proof}
(a) If $r\in S$, then $U(\hat r+Mu)=U(\hat r)+\frac32Mu$ by \cref{lem:steps}(f), and if $r\notin S$,
then $D(\hat r+Mu)=D(\hat r)+\frac12Mu$; and $\frac32Mu\equiv\frac12Mu\equiv\frac M2u\pmod M$.

(b) If $x\equiv x'\pmod{2^j}$ with $j\ge k+1$, then $x$ and $x'$ have the same residue $r$ modulo
$M$ and the same parity, so the same step is applied to both, and
$T_S(x)-T_S(x')=f(r)(x-x')\equiv0\pmod{2^{j-1}}$. By induction, $T_S^i(x)\equiv T_S^i(x')$ modulo
$2^{k+t-i}$, and in particular modulo $M$, for $i\le t$ if $x\equiv x'\pmod{2^{k+t}}$. So the
residue sequence depends only on $x\bmod2^{k+t}$. Now let $x\equiv x'\equiv r_0\pmod M$ with
$x\not\equiv x'\pmod{2^{k+t}}$, and let $k+i$, with $0\le i<t$, be the exact power of $2$ that
divides $x-x'$. As before, the residue sequences agree up to index $i$, the first $i$ steps are
the same, and $T_S^i(x)-T_S^i(x')=3^b(x-x')/2^i$ for some $b$; this difference is divisible by $M$
but not by $2M$. The same step is applied once more, so $T_S^{i+1}(x)-T_S^{i+1}(x')$ is divisible
by $M/2$ but not by $M$, and the residues at index $i+1$ differ. So distinct classes give distinct
sequences. By (a) every sequence is a path in $\Res_k(S)$, and there are exactly $2^t$ paths of
length $t$ from $r_0$, so the correspondence is a bijection.

(c) follows from (b).
\end{proof}

The lemma has a consequence for divergent orbits: they cannot be found by following a cycle of
the residue graph.

\begin{lemma}\label{lem:aperiodic}
Let $S$ be periodic modulo $2^k$. If the orbit of $x$ under $T_S$ diverges, then the sequence of
its residues modulo $2^k$ is not eventually periodic.
\end{lemma}

\begin{proof}
Suppose that the residues of $x_t=T_S^t(x)$ are periodic with period $\ell$ from $t_0$ on, and let
$y=x_{t_0}$ and $y'=x_{t_0+\ell}=T_S^\ell(y)$. Then $y$ and $y'$ have the same residue sequence, so
by \cref{lem:fresh}(b) they are congruent modulo $2^{k+t}$ for every $t$, and $y=y'$. So the orbit
is eventually periodic.
\end{proof}

\begin{remark}[Cycles of the residue graph as $2$-adic cycles]\label{rem:2adic}
The formulas $x\mapsto(3x+e)/2$ and $x\mapsto(x+e)/2$, with $e=x\bmod2$, also make sense for
$2$-adic integers $x$, and \cref{lem:fresh} holds for the resulting extension of $T_S$ to $\Z_2$.
Hence an infinite path in $\Res_k(S)$ is the residue sequence of exactly one $2$-adic integer. If
the path goes around a cycle of length $\ell$ with $a$ vertices in $S$ forever, this $2$-adic
integer $x$ satisfies $T_S^\ell(x)=x$. On the numbers that follow the cycle for $\ell$ steps,
$T_S^\ell$ is the affine map $y\mapsto(3^ay+d)/2^\ell$ for a nonnegative integer $d$, so
$x=d/(2^\ell-3^a)$ is a rational number. In particular, the cycles with $3^a>2^\ell$ are the
residues of periodic orbits of the extended map that consist of rational numbers $x\le0$, and
$x<0$ unless all vertices of the cycle are even residues. We use this in \cref{sec:collatz}.
\end{remark}

\subsection{The effective up-frequency}

\begin{definition}\label{def:pi}
Let $S$ be periodic modulo $2^k$. A \emph{closed class} of $\Res_k(S)$ is a strongly connected set
of vertices that no edge leaves. On a closed class $C$ the residue chain is irreducible and has a
unique stationary distribution $\mu_C$. The \emph{effective up-frequency} of $S$ on $C$ is
\[
\pi_C(S)=\sum_{r\in C\cap S}\mu_C(r).
\]
If $\Res_k(S)$ has a single closed class we write $\pi(S)$.
\end{definition}

In words: a closed class is a set of residues that the residue chain cannot leave once it has
entered it, and inside which it can get from every residue to every other one. The residues
outside all closed classes are visited only finitely often by the chain. The number $\pi_C(S)$
is the proportion of time that the chain spends at residues where the map goes up, once it has
settled in $C$.

By the ergodic theorem for finite Markov chains, the residue chain enters some closed class $C$
after finitely many steps almost surely, and then the proportion of its steps taken at vertices in
$S$ converges to $\pi_C(S)$. By \cref{lem:fresh}(c), $\pi_C(S)$ is therefore the proportion of
steps up of the orbit of a typical number whose residues end up in $C$, over any fixed number of
steps that is large compared with the time needed to enter $C$. The numbers $\pi_C(S)$ are
rational, so they are never equal to $\log_32$.

For the two copies of the Collatz map the uniform distribution is stationary, since every residue
has exactly two incoming edges, and $\pi(\odd)=\pi(\even)=1/2$, the density. In general $\pi$ and
the density are unrelated, and they can lie on opposite sides of $\log_32$; see \cref{tab:pi}. The
reason is the same as for arbitrary sets: what matters is not how many residues lie in $S$, but how
often the orbits visit them.

\begin{table}[ht]
\centering
\caption{Density and effective up-frequency $\pi$ of some periodic sets; each of these residue
graphs has a single closed class. Residues are taken modulo the given modulus, and $0$ stands for
the multiples of the modulus. In the last four rows $0.438=7/16$, $0.786=11/14$,
$0.688=11/16$, $0.303=10/33$ and $0.420=73/174$; the last set has the largest $\pi$ among the
uniformly descending sets modulo $16$ (\cref{tab:uniformcount}), and the one before it is the set
of \cref{thm:mod16}.}\label{tab:pi}
\small
\begin{tabular}{@{}lrlll@{}}
\toprule
$S$ & modulus & density & $\pi(S)$ & remark\\
\midrule
$\odd$, $\even$ & $2$ & $1/2$ & $1/2$ & the Collatz map\\
$4\N-3$ & $4$ & $1/4$ & $1/6$ & no two steps up in a row\\
$\N\setminus(4\N-1)$ & $4$ & $3/4$ & $5/6$ & $Z$-numbers (\cref{prop:mahler})\\
$\{1,2,7,9,10,11,15\}$ & $16$ & $0.438$ & $0.786$ & sparse but expanding\\
$\{0,1,2,4,7,8,9,12,13,14,15\}$ & $16$ & $0.688$ & $2/5$ & dense but contracting\\
$\odd$ without $9,15$ & $16$ & $3/8$ & $0.303$ & uniformly descending\\
$\{1,3,6,7,8,11,13\}$ & $16$ & $7/16$ & $0.420$ & uniformly descending\\
\bottomrule
\end{tabular}
\end{table}

The two middle examples of \cref{tab:pi} are the extreme cases modulo $16$: among all sets that
are periodic modulo $16$, they maximize $\pi-\text{density}$ subject to
$\text{density}<\log_32<\pi$, and $\text{density}-\pi$ subject to
$\pi<\log_32<\text{density}$. The effective up-frequency decides, as in the theorem of Terras and
Everett for the Collatz map \cite{Ter76,Eve77}, whether a typical orbit eventually falls below its
starting value:

\begin{theorem}\label{thm:terras}
Let $S$ be periodic modulo $2^k$.
\begin{enumerate}[label=(\alph*)]
\item If $\pi_C(S)<\log_32$ for every closed class $C$ of $\Res_k(S)$, then the set of $x$ with
$T_S^t(x)<x$ for some $t\ge1$ has density $1$.
\item If $\pi_C(S)>\log_32$ for some closed class $C$, then there are $c>0$ and $\lambda>1$ such
that for every large $t$ the set of $x$ with $T_S^t(x)\ge\lambda^tx$ has lower density at least
$c$.
\end{enumerate}
\end{theorem}

\begin{proof}
Let $a_t$ be the number of steps up among the first $t$ steps of the residue chain started from
the uniform distribution, that is, the number of the first $t$ vertices of its path that lie in
$S$. By the ergodic theorem, $a_t/t$ converges almost surely to $\pi_C(S)$, where $C$ is the closed
class that the chain enters.

(a) Choose $\varepsilon>0$ with $\pi_C(S)<\log_32-2\varepsilon$ for all $C$. Then
$P(a_t\ge(\log_32-\varepsilon)t)\to0$. Let $t$ be so large that $e\,3^{-\varepsilon t}<1$, and let
$x\ge t$ be a number whose residue sequence of length $t$ has fewer than $(\log_32-\varepsilon)t$
vertices in $S$. If $T_S^i(x)\ge x$ for all $i<t$, then by \cref{prop:critical}(a)
\[
T_S^t(x)\le x\,\frac{3^{a}}{2^t}\Bigl(1+\frac1x\Bigr)^t<x\cdot3^{-\varepsilon t}\cdot e<x.
\]
So every such $x$ falls below itself within $t$ steps. By \cref{lem:fresh}(c), the set of the
other numbers $x\ge t$ has upper density at most $P(a_t\ge(\log_32-\varepsilon)t)$, which tends to
$0$ as $t\to\infty$.

(b) Choose $\varepsilon>0$ with $\pi_C(S)>\log_32+2\varepsilon$. Started from the uniform
distribution on $C$, the chain stays in $C$, so $P(a_t\ge(\log_32+\varepsilon)t\mid r_0\in C)\to1$.
For every $x$ whose residue sequence of length $t$ has at least $(\log_32+\varepsilon)t$ vertices
in $S$, \cref{prop:critical}(a) gives $T_S^t(x)\ge x\,3^{a}/2^t\ge3^{\varepsilon t}x$. By
\cref{lem:fresh}(c), these $x$ form a union of residue classes of density at least
$\frac{\#C}{2^k}P(a_t\ge(\log_32+\varepsilon)t\mid r_0\in C)$, which is at least
$c=\#C/2^{k+1}$ for large $t$. Take $\lambda=3^\varepsilon$.
\end{proof}

Part (b) does not prove that $T_S$ is unbounded: an orbit may grow for a long time and still come
down. For the ``sparse but expanding'' set of \cref{tab:pi}, of density $7/16$, the residue graph
has a single closed class, and the proof of (b) shows that for every $\varepsilon>0$ and every
large $t$ all numbers $x$ outside a set of upper density $\varepsilon$ satisfy
$T_S^t(x)\ge1.18^t\,x$, since $3^{11/14}/2>1.18$; but we cannot prove that a single orbit diverges. By \cref{lem:aperiodic}, any divergent orbit would have
to visit the residues in an aperiodic way.

\subsection{Uniform descent}\label{sec:uniform}

\Cref{thm:terras}(a) only says that almost every number falls below itself. A much stronger
property can follow from the residue graph directly: that every long path in it descends.

\begin{definition}\label{def:uniform}
A set $S$ that is periodic modulo $2^k$ is \emph{uniformly descending} if every cycle of
$\Res_k(S)$ of length $\ell$ with $a$ vertices in $S$ satisfies $3^a<2^\ell$.
\end{definition}

Equivalently, the cycles of $\Res_k(S)$ all have a proportion $a/\ell<\log_32$ of vertices in $S$.
Since a closed walk is made of simple cycles, it suffices to check the simple cycles, and this can
be done in time polynomial in $2^k$ by computing the maximum mean weight of a cycle with weights
$\log f(r)$ (Karp's algorithm). The definition does not depend on $k$: a set that is periodic modulo
$2^k$ is also periodic modulo $2^{k+1}$, reduction modulo $2^k$ maps the cycles of
$\Res_{k+1}(S)$ to closed walks of $\Res_k(S)$ with the same steps, and conversely every cycle of
$\Res_k(S)$ is the reduction of the closed walk of $\Res_{k+1}(S)$ formed by the residues of the
corresponding $2$-adic periodic orbit (\cref{rem:2adic}). A uniformly descending set
has $\pi_C(S)<\log_32$ for all closed classes $C$: the stationary chain uses the edges of $C$ with
frequencies that form a circulation, a circulation is a positive combination of simple cycles, and
on each of these the proportion of vertices in $S$ is less than $\log_32$. The converse is false.
The effective up-frequency is an average over typical paths, while uniform descent is a condition
on the worst path: the Collatz set has $\pi=1/2<\log_32$, but it is not uniformly descending,
because of the loop at the residue $-1$ (\cref{ex:residue4}). The set $4\N-3$ of the same example
is uniformly descending, since at most half of the vertices of any of its cycles lie in $S$.

\begin{theorem}\label{thm:uniform}
Let $S$ be periodic modulo $M=2^k$. The following are equivalent.
\begin{enumerate}[label=(\roman*)]
\item $S$ is uniformly descending.
\item There are weights $c\colon\Z/M\to\N$ and a number $\theta<1$ such that
$f(r)\,c(s)\le\theta\,c(r)$ for every edge $r\to s$ of $\Res_k(S)$.
\item There is $t\ge1$ such that $3^a<2^t$ for every path of length $t$ in $\Res_k(S)$ with $a$
vertices in $S$ (not counting its last vertex).
\item There is $t\ge1$ such that $T_S^t(x)<x$ for all sufficiently large $x$.
\end{enumerate}
\end{theorem}

\begin{corollary}[Certificates]\label{cor:certificate}
Let $c$ and $\theta$ be as in (ii), let $x_1$ be the least integer with $\theta(1+1/x_1)<1$, and let
$B=\floor{c_{\max}U(x_1)/c_{\min}}+1$, where $c_{\max}$ and $c_{\min}$ are the largest and the
smallest weight. Then every orbit of $T_S$ is eventually smaller than $B$. In particular $T_S$ is
bounded, all its cycles lie in $[1,B]$, and if the orbits of all $x<B$ end in a given set of
cycles, then so do all orbits.
\end{corollary}

Thus a uniformly descending set comes with a proof that can be checked by hand: a table of
$2^k$ weights, the inequalities on the $2^{k+1}$ edges, and a finite computation below $B$. The
weights play the role of a Lyapunov function $\Phi(x)=c(x\bmod M)\,x$ that decreases at every
step. The size $x$ itself does not decrease at every step, of course. But a residue from which
steps up are still to come gets a larger weight, so that the weighted size $\Phi$ has already
``paid'' for these steps, and each single step makes it smaller.

Here is the smallest example, for $S=4\N-3$ and the modulus $4$. Let $\Phi(x)=7x$ if
$x\equiv1\pmod4$ and $\Phi(x)=4x$ otherwise. A number $x\equiv1\pmod4$ goes up to the even number
$(3x+1)/2$, so $\Phi$ changes from $7x$ to $6x+2$, which is smaller for $x>2$. Any other number
$x$ goes down to at most $(x+1)/2$, so $\Phi$ changes from $4x$ to at most $\frac72(x+1)$, which
is smaller for $x>7$. So $\Phi$ decreases at every step from a number $x\ge8$, and every orbit
comes below $8$. From there it cannot climb back above $U(5)=8$, and the numbers up to $8$ all
reach the cycle $(1,2)$. This proves that every orbit of $T_{4\N-3}$ reaches $(1,2)$. In the
notation of \cref{thm:uniform}(ii) the weights are $c=(4,7,4,4)$ at the residues $0,1,2,3$, and
$\theta=7/8$. (For this set the result also follows from \cref{prop:runs}(b); the point is that
the same kind of proof works for sets with longer runs, such as the set of \cref{thm:mod16}.)

\begin{proof}[Proof of \cref{thm:uniform}]
Let $w(r)=\log f(r)$, and define the weight of a path as the sum of $w$ over its vertices other
than the last one; a path of length $t$ with $a$ such vertices in $S$ has weight
$\log(3^a/2^t)$.

(i)$\Rightarrow$(iii). Every closed walk splits into simple cycles, and there are finitely many
simple cycles, each of negative weight; let $-\gamma<0$ be the largest of their weights divided
by their lengths. A path of length $t$ splits into a simple path of length less than $M$ and
simple cycles of total length more than $t-M$, so its weight is less than
$M\log\frac32-\gamma(t-M)$, which is negative for large $t$.

(iii)$\Rightarrow$(iv). Let $\eta<1$ be the largest value of $3^a/2^t$ over all paths of length
$t$. By \cref{lem:fresh}(a), the residues of an orbit form a path, and $T_S^i(x)\ge x/2^i$, since
every step at least halves. So \cref{prop:critical}(a) gives
$T_S^t(x)\le x\,\eta\,(1+2^t/x)^t$, which is less than $x$ for large $x$.

(iv)$\Rightarrow$(i). Suppose that a cycle $r_0\to r_1\to\dots\to r_\ell=r_0$ has $3^a>2^\ell$.
Its weights $w(r_0),\dots,w(r_{\ell-1})$ have a positive sum, so for a suitable rotation all
partial sums are nonnegative: start at a vertex $r_m$, $0\le m<\ell$, for which the partial sum
$\sum_{i<m}w(r_i)$ is smallest. We may therefore assume that the path $P$ that goes around the
cycle from $r_0$ repeatedly has weight at least $0$ after any number of steps. Given $t$, by
\cref{lem:fresh}(b) there are arbitrarily large $x$ whose residue sequence begins with the first
$t$ steps of $P$, and for them $T_S^t(x)\ge x\,3^a/2^t\ge x$ by \cref{prop:critical}(a), where $a$
now counts the steps up among the first $t$. So (iv) fails for every $t$.

(i)$\Rightarrow$(ii). Let $\mu^*<0$ be the largest mean weight of a simple cycle, choose
$\mu\in(\mu^*,0)$, and let $\psi(r)$ be the supremum of $\sum_{i<j}(w(r_i)-\mu)$ over all paths
$r=r_0\to\dots\to r_j$, $j\ge0$. Every cycle has negative weight for the weights $w-\mu$, so
$0\le\psi(r)\le M\max_r|w(r)-\mu|$, and $\psi(r)\ge w(r)-\mu+\psi(s)$ for every edge $r\to s$. With
$c(r)=e^{\psi(r)}$ and $\theta=e^\mu$ this is $f(r)c(s)\le\theta c(r)$. Replacing $\theta$ by a
slightly larger number below $1$, we may perturb the weights to rational numbers and then multiply
them by a common denominator.

(ii)$\Rightarrow$(i). Along a cycle, $3^a/2^\ell=\prod_if(r_i)\le\prod_i\theta
c(r_i)/c(r_{i+1})=\theta^\ell<1$.
\end{proof}

\begin{proof}[Proof of \cref{cor:certificate}]
Let $\Phi(x)=c(x\bmod M)\,x$, a positive integer. For $x\equiv r\pmod M$ we have
$T_S(x)\le f(r)\,x\,(1+1/x)$, and $T_S(x)\bmod M$ is a successor $s$ of $r$, so
\[
\Phi(T_S(x))=c(s)\,T_S(x)\le\frac{\theta\,c(r)}{f(r)}\,f(r)\,x\Bigl(1+\frac1x\Bigr)
=\theta\Bigl(1+\frac1x\Bigr)\Phi(x).
\]
Hence $\Phi(T_S(x))<\Phi(x)$ for $x\ge x_1$. Since $\Phi$ takes positive integer values, no orbit
stays in $[x_1,\infty)$ forever. Whenever an orbit passes from $[1,x_1)$ to $[x_1,\infty)$, it
arrives at a number $y\le U(x_1)$, and from then on, as long as it stays in $[x_1,\infty)$, its
elements $z$ satisfy $c_{\min}z\le\Phi(z)\le\Phi(y)\le c_{\max}U(x_1)$, so $z<B$. So after its first
visit to $[1,x_1)$ the orbit stays below $B$. The remaining claims follow.
\end{proof}

We checked all sets that are periodic modulo $4$, $8$ and $16$; \cref{tab:uniformcount} lists how
many of them are uniformly descending, and the largest effective up-frequency and density among
these. Uniform descent forces $\pi$ well below $\log_32$, at least for small moduli. We do not
know whether uniformly descending sets with $\pi$ arbitrarily close to $\log_32$ exist for larger
moduli (\cref{q:supremum}).

\begin{table}[ht]
\centering
\caption{The numbers of all sets and of the uniformly descending sets that are periodic modulo
$2^k$, $k\le4$, and the largest effective up-frequency and density among the latter.}
\label{tab:uniformcount}
\small
\begin{tabular}{@{}rrrrlrl@{}}
\toprule
modulus & sets & descending & \multicolumn{2}{l}{largest $\pi$, attained by} & \multicolumn{2}{l@{}}{largest density, attained by}\\
\midrule
$4$ & $16$ & $3$ & $1/4$ & $\{2\}$ & $1/4$ & $\{1\}$\\
$8$ & $256$ & $20$ & $5/18$ & $\{3,4,6\}$ & $3/8$ & $\{1,2,5\}$\\
$16$ & $65{,}536$ & $1{,}850$ & $73/174$ & $\{1,3,6,7,8,11,13\}$ & $1/2$ & $\{1,2,4,5,7,8,9,13\}$\\
\bottomrule
\end{tabular}
\end{table}

\subsection{Mahler's \texorpdfstring{$Z$}{Z}-numbers}

Mahler \cite{Mah68} called a real number $\xi>0$ a \emph{$Z$-number} if
$0\le\{\xi(3/2)^n\}<1/2$ for all $n\ge0$, and asked whether $Z$-numbers exist; he proved that
there is at most one in every interval $[m,m+1)$, and that the number of $m\le X$ for which there
is one is $O(X^{0.7})$. The question is still open; see also Flatto, Lagarias and Pollington
\cite{FLP95}. It turns out to be a question about one map of our family, or more precisely, it is
implied by a question about one map.

\begin{proposition}\label{prop:mahler}
Let $\xi$ be a $Z$-number and $g_n=\floor{\xi(3/2)^n}$. Then $g_{n+1}=U(g_n)$ and
$g_n\not\equiv3\pmod4$ for all $n$ with $g_n\ge1$. Consequently, the set $S=\N\setminus(4\N-1)$
contains a tail of a chain, and $T_S$ has an increasing orbit.
\end{proposition}

\begin{proof}
Write $\xi(3/2)^n=g_n+f_n$ with $0\le f_n<1/2$. If $g_n$ is even, then
$\xi(3/2)^{n+1}=\frac32g_n+\frac32f_n$ with $\frac32g_n\in\Z$ and $\frac32f_n<\frac34$, so
$g_{n+1}=\frac32g_n=U(g_n)$. If $g_n$ is odd, then
$\xi(3/2)^{n+1}=\frac{3g_n+1}2+\frac{3f_n-1}2$, and $-\frac12\le\frac{3f_n-1}2<\frac14$. If
$3f_n<1$, the fractional part $f_{n+1}=\frac{3f_n+1}2$ would be at least $\frac12$; so
$3f_n\ge1$, $g_{n+1}=\frac{3g_n+1}2=U(g_n)$, and $f_{n+1}=\frac{3f_n-1}2<\frac14$. Since
$f_{n+1}<\frac13$, the same argument applied to $n+1$ shows that $g_{n+1}$ cannot be odd. So every
odd $g_n$ has $\frac{3g_n+1}2$ even, that is, $g_n\equiv1\pmod4$. As $g_n\to\infty$, the numbers
$g_n\ge1$ form a tail of a chain that avoids $4\N-1$, and under $T_S$ the orbit of its first
element increases.
\end{proof}

\begin{corollary}\label{cor:mahler}
If every chain contains infinitely many numbers $\equiv3\pmod4$, that is, if $T_S$ has no
increasing orbit for $S=\N\setminus(4\N-1)$, then there are no $Z$-numbers.
\end{corollary}

The set $\N\setminus(4\N-1)$ is periodic modulo $4$ with $\pi=5/6>\log_32$ (\cref{tab:pi}), so by
\cref{thm:terras}(b) a positive proportion of all orbits grows exponentially for a long time, and
we expect $T_S$ to have many divergent orbits. The corollary asks for much less than boundedness:
only that no orbit avoids the residue class $3$ modulo $4$ forever while going up at every step.
In the other direction, let $e_m$ be the parity of $U^m(c)$. Since
$(2/3)^{m+1}U^{m+1}(c)=(2/3)^mU^m(c)+\frac13(2/3)^me_m$, summing over $m\ge n$ gives
$K_c(3/2)^n=U^n(c)+\frac13\sum_{j\ge0}(\frac23)^je_{n+j}$, so a chain tail from $c$ that
avoids $4\N-1$, and hence has no two consecutive odd elements, gives a number $\xi=K_c$ with
$0<\{\xi(3/2)^n\}\le\frac13(1+\frac49+\frac{16}{81}+\cdots)=\frac35$ for all $n\ge0$, but not
necessarily a $Z$-number. So the question of \cref{cor:mahler} lies between Mahler's question and
the same question for the interval $(0,3/5]$ in place of $[0,1/2)$.

\section{How far is the Collatz map from provable behaviour?}\label{sec:collatz}

The Collatz map is $T_S$ for $S=\odd$. We now ask which modifications of this set make the
behaviour of the map provable, in either direction: that every orbit reaches the cycle $(1,2)$, or
that some orbit diverges. A modification is described by the set $S\,\triangle\,(\odd)$ of changed
numbers, which consists of the removed odd numbers $(\odd)\setminus S$ and the added even numbers
$S\cap\even$, and we measure its size by the number of changed numbers below $N$, or by its
density. This gives a notion of distance between the Collatz map and the maps whose behaviour we
can prove. The answers are very different in the two directions (\cref{tab:distance}).

\subsection{Finite changes}

By \cref{lem:finite}, changing $\odd$ at finitely many numbers does not change whether there are
divergent orbits:

\begin{corollary}\label{cor:finite}
If $S\,\triangle\,(\odd)$ is finite, then $T_S$ has a divergent orbit if and only if the Collatz
map has one.
\end{corollary}

So every modification that makes the existence or the absence of divergent orbits provable must be
infinite, unless it settles the same question for the Collatz map itself. We look for infinite
modifications that are as thin as possible.

\subsection{Divergence from a logarithmically thin addition}

The cheapest way to create a divergent orbit is to complete a chain: an orbit goes up forever if
and only if it runs along a chain inside $S$ (\cref{prop:runs}(a)), and the odd elements of every
chain are already in $\odd$.

\begin{proposition}\label{prop:mahleradd}
Let $A$ be the set of even elements of the chain $\Ch_1=\{U^r(1):r\ge0\}$,
\[
A=\{2,8,12,18,62,140,210,710,1598,3596,5394,18206,40964,61446,\dots\}.
\]
Then the orbit of $1$ under $T_{(\odd)\cup A}$ is $\Ch_1$ and increases, and
$\#(A\cap[1,N])\le\log_{3/2}N$ for all $N\ge1$. The same holds for the even elements of any chain.
\end{proposition}

\begin{proof}
Every element of $\Ch_1$ is odd or lies in $A$, so $\Ch_1\subseteq(\odd)\cup A$, and the orbit of
$1$ goes up at every step. By \cref{lem:steps}(d), $U^r(1)>K_1(3/2)^r-1\ge(3/2)^r$ for $r\ge2$, and
also for $r=0,1$, so $\Ch_1$ has at most $\log_{3/2}N+1$ elements in $[1,N]$, one of which is the
odd number $1$.
\end{proof}

Up to $10^{60}$ the chain $\Ch_1$ has $340$ elements, of which $173$ are even, so the added set
has about $\frac12\log_{3/2}N$ elements below $N$. The parities of the numbers $\floor{K_1(3/2)^n}$
are the subject of Mahler's problem and are not understood, but they look random, and this count
is what one expects. For instance, only $18$ even numbers below $10^6$ are added.

The effect of this thin addition is drastic. Under the modified map every element of $\Ch_1$ goes
up, so an orbit that meets $\Ch_1$ increases from then on. An orbit of the modified map agrees
with the Collatz orbit of the same number until that orbit first meets $\Ch_1$, and a Collatz
orbit that reaches $1$ meets $\Ch_1$, since $1\in\Ch_1$. Hence, if the Collatz conjecture is true,
then \emph{every} orbit of $T_{(\odd)\cup A}$ diverges.

Could fewer additions suffice, say $o(\log N)$ below $N$? By \cref{prop:critical}(b) a divergent
orbit goes up a proportion at least $\log_32>1/2$ of the time, while the steps up of a typical
orbit at odd numbers make up only about half of its steps; so if the orbit is typical, it needs
steps up at added even numbers at a positive proportion of its steps, and since it grows
exponentially, this means order $\log N$ added numbers below $N$. An atypical orbit that visits odd
numbers unusually often might need fewer.

\begin{remark}[A first-moment heuristic]\label{rem:heuristic}
Model the parities of the numbers visited by the paths from a given number by independent fair
coins, and count the paths of length $t$ in $\Gp{\odd}{\N}$ (up at odd numbers, free at even
numbers) that visit about $ot$ odd numbers and go up at about $\kappa t$ even numbers. Their
expected number is $\exp\bigl(t(H(o)+(1-o)H(\frac{\kappa}{1-o})-\log2)+o(t)\bigr)$, where $H$ is
the entropy function in natural logarithms, and such a path grows by the factor
$\exp\bigl(t((o+\kappa)\log3-\log2)\bigr)$. Minimizing the number of additions per unit of
logarithmic growth, $\kappa/((o+\kappa)\log3-\log2)$, subject to a nonnegative first exponent gives
about $0.23$, at $o\approx0.69$ and $\kappa\approx0.021$, compared with $1/\log\frac94\approx1.23$
for the even elements of a chain. So the model suggests that order $\log N$ additions below $N$ are
needed, but with a smaller constant than in \cref{prop:mahleradd}: as $\kappa\to0$ the constraint
forces $o\to1/2$ and negative growth.
\end{remark}

\subsection{Convergence from removing residue classes}

In order to prove that every orbit converges, we must control all orbits at once; the tool is uniform
descent (\cref{sec:uniform}). The Collatz set itself is not uniformly descending: its residue
graph modulo $2^k$ has the loop $2^k-1\to2^k-1$ at a vertex in $S$, the shadow of the fixed point
$-1$ of the Collatz map on the $2$-adic integers, since $U(2^k-1)=3\cdot2^{k-1}-1\equiv-1+2^{k-1}$.
Removing residue classes from $\odd$ can break such cycles. The smallest modulus at which a removal
of density less than $1/4$ works is $16$:

\begin{theorem}\label{thm:mod16}
Let $S$ be the set of odd numbers $x$ with $x\not\equiv9,15\pmod{16}$. Then every orbit of $T_S$
reaches the cycle $(1,2)$.
\end{theorem}

Explicitly, the map of the theorem is
\[
x\mapsto\begin{cases}
x/2, & x\text{ even},\\
(x+1)/2, & x\equiv9\text{ or }15\pmod{16},\\
(3x+1)/2, & \text{all other odd }x.
\end{cases}
\]
So turning the steps at the odd numbers $\equiv9$ and $\equiv15\pmod{16}$, an eighth of all
positive integers, from up into down gives a Collatz-type map whose convergence can be proved in a
few lines. For example, the Collatz orbit of $27$ needs $70$ steps to reach $1$ and climbs to
$4616$ on the way; under the modified map the orbit is $27,41,21,32,16,8,4,2,1$, because
$41\equiv9\pmod{16}$ goes down to $21$ instead of up to $62$.

\begin{table}[ht]
\centering
\caption{A certificate for \cref{thm:mod16}: the residue graph of $S$ modulo $16$ with weights
$c(r)$. The last column is the largest ratio $f(r)c(s)/c(r)$ over the two successors $s$; all are
at most $\theta=75/76$.}\label{tab:cert16}
\begin{tabular}{@{}rclrr@{\qquad}rclrr@{}}
\toprule
$r$ & step & successors & $c(r)$ & ratio & $r$ & step & successors & $c(r)$ & ratio\\
\midrule
$0$ & down & $8,0$ & $16$ & $1/2$ & $8$ & down & $4,12$ & $16$ & $1/2$\\
$1$ & up & $2,10$ & $25$ & $24/25$ & $9$ & down & $5,13$ & $16$ & $25/32$\\
$2$ & down & $1,9$ & $16$ & $25/32$ & $10$ & down & $5,13$ & $16$ & $25/32$\\
$3$ & up & $5,13$ & $38$ & $75/76$ & $11$ & up & $1,9$ & $38$ & $75/76$\\
$4$ & down & $2,10$ & $16$ & $1/2$ & $12$ & down & $6,14$ & $16$ & $15/16$\\
$5$ & up & $8,0$ & $25$ & $24/25$ & $13$ & up & $4,12$ & $25$ & $24/25$\\
$6$ & down & $3,11$ & $20$ & $19/20$ & $14$ & down & $7,15$ & $30$ & $29/30$\\
$7$ & up & $11,3$ & $58$ & $57/58$ & $15$ & down & $8,0$ & $16$ & $1/2$\\
\bottomrule
\end{tabular}
\end{table}

\begin{proof}
The proof is an instance of \cref{cor:certificate}: a weighted size $\Phi(x)=c(x\bmod16)\,x$
decreases at every step from every large enough $x$, and the remaining small numbers are checked
directly. $S$ is periodic modulo $16$, and \cref{tab:cert16} lists its residue graph: for instance $7\in S$
goes up to $U(7)=11$, so its successors are $11$ and $11+8\equiv3$, while $9\notin S$ goes down to
$D(9)=5$, with successors $5$ and $13$. The weights satisfy $f(r)c(s)\le\frac{75}{76}c(r)$ on all
$32$ edges, as the last column shows. In \cref{cor:certificate} we have $\theta=75/76$, so
$x_1=76$, $U(76)=114$, and $B=\floor{58\cdot114/16}+1=414$. Concretely, $\Phi(x)=c(x\bmod16)\,x$
decreases at every step from a number $x\ge76$, and every orbit eventually stays below $414$. A
computation shows that every $x$ with $2\le x\le414$ falls below itself within at most $8$ steps,
so by induction on $x$ every $x\le414$ reaches $1$. Since $1\in S$, the orbit of $1$ is the cycle
$(1,2)$.
\end{proof}

Why the classes $9$ and $15$? Call a cycle of a residue graph \emph{expanding} if $3^a>2^\ell$,
that is, if going around it multiplies a number by more than $1$; a set is uniformly descending
exactly when its residue graph has no expanding cycle. The residue graph of $\odd$ modulo $16$
has exactly eleven expanding simple cycles. The shortest are the loop at $15$, the cycle
$6\to11\to9\to6$ and the cycle $7\to11\to9\to14\to7$. A modification can make the residue graph
uniformly descending only if it changes the step at some residue on each of these eleven cycles.
Nine of them pass through the residue $9$, and the same nine pass through the residue $11$; the
other two, among them the loop, pass through $15$. So with two residues there are exactly two
candidates, $\{9,15\}$ and $\{11,15\}$, and both work: the second choice gives a theorem like
\cref{thm:mod16} as well, with $\pi=13/48$. Changing the steps at even residues gives no further
possibilities with two residues.

\begin{remark}[Shadows of negative cycles]\label{rem:shadows}
By \cref{rem:2adic}, the cycles of the residue graph of $\odd$ with $3^a>2^\ell$ are the residues
of periodic orbits of the Collatz map $T$, extended to the $2$-adic integers, that consist of
negative rational numbers. Modulo $16$, the loop at $15$ is the fixed point $-1$, the cycle
$6\to11\to9\to6$ is the cycle $-10\to-5\to-7\to-10$ of $T$ on the negative integers, and the
cycle $7\to11\to9\to14\to7$ is the cycle
$-\frac{19}{11}\to-\frac{23}{11}\to-\frac{29}{11}\to-\frac{38}{11}\to-\frac{19}{11}$. Of the three
known cycles of $T$ on the negative integers, the third, of length $11$ through $-17$, passes
through $-17\equiv15$ as well.
So to make a periodic modification of the Collatz map uniformly descending, one has to destroy all
the negative rational cycles of the extended map that are visible modulo $2^k$. Finding the
smallest such removal is a hitting-set problem for the cycles of a finite graph.
\end{remark}

For every $k$ let $\delta_k$ be the smallest density of $S\,\triangle\,(\odd)$ over all sets $S$
that are periodic modulo $2^k$ and uniformly descending, and let $\delta_k^-\ge\delta_k$ be the same
minimum over such sets $S\subseteq\odd$, that is, over removals only. Both sequences are
non-increasing in $k$, since uniform descent does not depend on the modulus. A modification has to
change at least one residue on every cycle of $\Res_k(\odd)$ with $3^a>2^\ell$, so $p$ pairwise
disjoint such cycles give $\delta_k\ge p/2^k$. On the other side, every removal for which the
residue graph is uniformly descending gives an upper bound, which becomes a theorem like
\cref{thm:mod16} once a certificate has been computed and the orbits below the bound $B$ of
\cref{cor:certificate} have been checked. \Cref{tab:deltak} shows what we found.

\begin{table}[ht]
\centering
\caption{Smallest periodic modifications of $\odd$ that are uniformly descending. Bold entries are
exact. $R_k$ is a set of removed odd residues of density $\#R_k/2^k$ that attains the upper bound
(listed in \cref{sec:computations}); for it, $\pi$ is the effective up-frequency of
$(\odd)\setminus R_k$, ``worst cycle'' the largest proportion $a/\ell$ over the cycles of its
residue graph, and $\theta$ and $B$ describe the certificate. An entry $[p,q]/2^k$ means that the
value lies between $p/2^k$ and $q/2^k$. For every $R_k$, all orbits reach $(1,2)$.}\label{tab:deltak}
\begin{tabular}{@{}rrlllllrr@{}}
\toprule
$k$ & $2^k$ & $\delta_k$ & $\delta^-_k$ & $\#R_k$ & $\pi$ & worst cycle & $\theta$ & $B$\\
\midrule
$2$ & $4$ & $\mathbf{1/4}$ & $\mathbf{1/4}$ & $1$ & $1/6$ & $1/2$ & $0.885$ & $21$\\
$3$ & $8$ & $\mathbf{1/4}$ & $\mathbf{1/4}$ & $2$ & $1/6$ & $1/2$ & $0.885$ & $21$\\
$4$ & $16$ & $\mathbf{1/8}$ & $\mathbf{1/8}$ & $2$ & $10/33$ & $3/5$ & $75/76$ & $414$\\
$5$ & $32$ & $\mathbf{1/8}$ & $\mathbf{1/8}$ & $4$ & $0.288$ & $3/5$ & $0.9718$ & $370$\\
$6$ & $64$ & $\mathbf{5/64}$ & $\mathbf{5/64}$ & $5$ & $0.322$ & $5/8$ & $0.9945$ & $1{,}068$\\
$7$ & $128$ & $\mathbf{8/128}$ & $\mathbf{8/128}$ & $8$ & $0.369$ & $5/8$ & $0.9945$ & $2{,}107$\\
$8$ & $256$ & $[10,12]/256$ & $\le12/256$ & $12$ & $0.364$ & $5/8$ & $0.9945$ & $1{,}611$\\
$9$ & $512$ & $[14,22]/512$ & $\le22/512$ & $22$ & $0.373$ & $5/8$ & $0.9945$ & $4{,}191$\\
$10$ & $1024$ & $[23,44]/1024$ & $\le44/1024$ & $44$ & $0.382$ & $17/27$ & $0.9988$ & $19{,}662$\\
\bottomrule
\end{tabular}
\end{table}

For $k=2$ and $k=3$ the best removal is the class $3\pmod4$, which leaves $4\N-3$ and is covered
by \cref{prop:runs}(b) as well. The proportion of changed numbers decreases slowly: $1/4$, $1/4$,
$1/8$, $1/8$, $5/64$, $1/16$, and then at most $0.047$, $0.043$ and $0.043$. As the modulus grows,
the surviving cycles come closer to the threshold ($a/\ell=3/5$, $5/8$, $17/27$, three fractions
just below $\log_32$), the certificates get tighter ($\theta\to1$), and the bounds $B$ for the finite checks tend to
grow.
Two features are worth noting. First, for $k\ge4$ removing classes lowers the effective
up-frequency from $1/2$ to between $0.29$ and $0.38$, far below $\log_32$, although only a few percent of all numbers
are changed: the removed classes are exactly the ones through which orbits climb. Second, for
$k\le7$ allowing changes at even residues as well does not help, $\delta_k=\delta_k^-$; we do not
know whether this persists.

\begin{remark}[Consequences for the Collatz map]\label{rem:sufficient}
Let $R$ be the set of numbers removed in \cref{thm:mod16} or in a row of \cref{tab:deltak}, and
let $S=(\odd)\setminus R$. The maps $T_S$ and $T$ differ only at the numbers of $R$, so an orbit
of the Collatz map $T$ is an orbit of $T_S$ until it meets $R$. Since every orbit of $T_S$ reaches
$1$, \emph{every orbit of the Collatz map reaches $1$ or contains a number of $R$}. Applied to
the tails of an orbit, this shows that a divergent orbit of $T$ would contain infinitely many
numbers of $R$, and that a cycle of $T$ other than $(1,2)$ would contain at least one. So the
Collatz conjecture is equivalent to the statement that every $x\equiv9$ or $15\pmod{16}$ reaches
$1$, and equally to the same statement for the $22$ classes of $R_9$ modulo $512$, which contain
$4.3\%$ of all numbers.

Reductions of this kind are known. A set is called \emph{sufficient} if the orbit of every positive
integer meets the orbit of some element of the set, and Monks \cite{Mon06} proved that every
arithmetic progression is sufficient. The property above is stronger, because the orbit itself has
to pass through the set. Monks, Monks, Monks and Monks \cite{MMMM13} call a set \emph{forward
sufficient} if every divergent orbit of $T$ contains an element of it. They prove that a set of
residues modulo $d$ is forward sufficient if every cycle of the residue graph of $T$ that avoids
these residues has a proportion of steps up below $\log_32$, and their examples, found by computer,
include the pairs $\{9,15\}$ and $\{11,15\}$ modulo $16$ of this section. Their criterion asks
less than uniform descent of $T_S$: only the cycles that avoid the chosen residues matter, and the
residues may be even. For $k\le7$ it still does not give smaller sets, because no set of fewer
than $2^k\delta_k$ residues modulo $2^k$ meets every expanding cycle of $\Res_k(\odd)$
(\cref{sec:computations}). What \cref{thm:mod16} and \cref{tab:deltak} add is the statement for
cycles, which comes from the same finite check, the minimality for $k\le7$, and the sets for the
moduli up to $1024$.

None of this is a shortcut for the numerical verification of the Collatz conjecture. The first
number of $R$ on the orbit of $x$ can be larger than $x$: from $27$ the orbit goes to $41$, and
from $39$ it reaches $89$ before any other number of $R$. A certificate only shows that this
number is less than $x\,c_{\max}/c_{\min}$ for $x\ge B$. So in order to verify the conjecture
for all $x\le N$ one would have to follow the numbers of $R$ up to a multiple of $N$, each of them
all the way to $1$. The sieves used in practice \cite{Bar25} discard the residue classes modulo
$2^k$ whose numbers fall below themselves within $k$ steps. The classes that are kept also meet
every orbit that does not reach $1$, a kept number has to be followed only until it falls below
itself, and $k$ can be taken large. At equal modulus the sets $R_k$ are smaller, except for
$k=5$, with $2,4,5,8,12,22,44$ classes for $k=4,\dots,10$ against $3,4,8,13,19,38,64$ kept
classes, but this does not outweigh the two disadvantages.
\end{remark}

\subsection{Boundedness from barriers}

If we only ask for boundedness, and not for convergence to $(1,2)$, there is a cheaper but
irregular way: create barriers.

\begin{proposition}\label{prop:barrierremove}
Let $B_1<B_2<\cdots$ be integers with $B_{i+1}\ge2^{B_i}$, and let $R$ be the set of odd numbers
in $\bigcup_i(2B_i/3,B_i]$. Then $T_S$ is bounded for $S=(\odd)\setminus R$, and $R$ has lower
density $0$. On the other hand, every $R\subseteq\odd$ for which $(\odd)\setminus R$ is trivially
bounded has upper density at least $1/6$.
\end{proposition}

\begin{proof}
Every $B_i$ is a barrier of $S$, so $S$ is trivially bounded. For $N=\floor{2B_{i+1}/3}$ we have
$R\cap[1,N]\subseteq[1,B_i]$, so $\#(R\cap[1,N])/N\le B_i/N\to0$. Conversely, if $B$ is a barrier
of $(\odd)\setminus R$, then $R$ contains all odd numbers in $(2B/3,B]$, at least $B/6-1$ of them,
so $\#(R\cap[1,B])\ge B/6-1$ for infinitely many $B$.
\end{proof}

Such a map is bounded, but we cannot say which cycles it has: below each barrier it is still a
Collatz map with some steps changed. Barriers also need a positive upper density, so in terms of
density they are no better than the periodic removals of \cref{tab:deltak}.

\subsection{The asymmetry}\label{sec:asymmetry}

The two directions behave very differently. A divergent orbit is a single path that never comes back down, and a set that contains it can be extremely thin: it only has to supply the steps up that the Collatz map does not already make, and along a chain these are the even elements. A proof of convergence must control every orbit, and for a periodic modification this means breaking every expanding cycle of the residue graph, that is, every negative rational cycle of the extended map
that is visible modulo $2^k$ (\cref{rem:shadows}). Conversely, for a periodic set a divergent orbit cannot be found by following a cycle of the residue graph: by \cref{lem:aperiodic} a divergent
orbit has aperiodic residues. The chains of \cref{prop:mahleradd} and the rays of
\cref{thm:barely} use the archimedean size of the numbers instead. Residues alone can prove
divergence when they force every orbit upward: for $S=\N\setminus(8\N-1)$ a step down at
$x\equiv7\pmod8$ leads to $(x+1)/2\equiv0\pmod4$ and is followed by two steps up, to $3(x+1)/4$
and to $9(x+1)/8$, so every orbit diverges.

Measured by counting functions, the Collatz set is therefore within $O(\log N)$ of a provably
divergent map, and within $0.043N$ of a map that provably converges. We do not know whether the
second distance can be made $o(N)$, whether removals alone, without additions, can produce a
divergent orbit, or whether $o(\log N)$ additions can (\cref{q:deltak,q:removal,q:addition}).

\section{Questions}\label{sec:questions}

Some of the questions that arose, roughly in the order of the sections.

\subsection*{Cycles}

\begin{question}\label{q:ratio}
Does every cycle satisfy $2^\ell<3^{a+1}$, that is, $a=\floor{\ell\log_32}$? This holds for
cycles with a single run of steps up (\cref{thm:G}(d)), for all cycles of length $\ell\le30$, and
for all $1{,}133{,}265{,}572$ cycles whose elements are all at most $160$; among the latter, the
largest ratio $2^\ell/3^a$ is $2.398$, for $\ell=71$. If the answer is negative, what is the
supremum of $2^\ell/3^a$?
\end{question}

\begin{question}\label{q:lengths}
Which lengths occur? Up to $30$ they are $1,2,5,8,10,13,15,16,21,24,26,27,29$, and occurrence is
not monotone in $2^\ell/3^{\floor{\ell\log_32}}$: there is no cycle of length $7$, with ratio
$1.580$, but there are $12$ of length $15$, with ratio $1.665$. Is every large $\ell$ a cycle
length, and how fast does the number of cycles of length $\ell$ grow?
\end{question}

\begin{question}\label{q:types}
Is there a map with infinitely many cycles and a divergent orbit (\cref{rem:types})?
\end{question}

\begin{question}\label{q:hamilton}
Is there a set $S$ such that the orbit of $1$ under $T_S$ contains every positive integer, that
is, does $G$ have a ray through all vertices? Such a ray has to start at $1$, since the only edges
into $1$ come from $1$ and $2$, and it has to continue with $2,3,5,8$, since $D(2)=1$, $D(3)=2$
and $D(5)=3$ have already been visited.
\end{question}

\subsection*{Families}

\begin{question}\label{q:hereditary}
Which sets $B$ are force-bounded, that is, for which $B$ has $\Gp\emptyset B$ no ray?
Trivially bounded sets and sets with $\run(B)\le1$ are, and sets that contain a tail of a chain
are not. The two copies of the Collatz map behave very differently here. For $B=\even$, every
$x\le20{,}000$ reaches at most $2{,}466$ vertices of $\Gp\emptyset{\even}$, and every cycle of a
map $T_S$ with $S\subseteq\even$ and all elements below $4\cdot10^6$ lies in $\{1\}$, in $\{2,3\}$,
or in one strongly connected set of $74$ vertices between $28$ and $4617$; this set contains the
Collatz orbit of $27$, shifted by one, up to its maximum. (For instance, following the Collatz orbit
of $27$ up to $445$, and then taking $(y-1)/2$ instead of $(3y+1)/2$ at $y=445$, $111$ and $55$,
returns to $27$.) For $B=\odd$, the strongly connected component of $15$ in the subgraph of
$\Gp\emptyset{\odd}$ induced by $[1,2^{28}]$ contains $19.6\%$ of all numbers in this interval. Is
$\even$ force-bounded? This would imply that no Collatz orbit diverges.
\end{question}

\begin{question}\label{q:removal}
Is there a set $S$ of odd numbers for which $T_S$ has a divergent orbit? In other words, can
removing odd numbers from the Collatz set, without adding even ones, produce divergence?
\end{question}

\begin{question}\label{q:forcing}
Does every force-divergent set, that is, every $A$ such that $T_S$ is unbounded for all $S\supseteq A$,
contain a tail of a chain?
\end{question}

\subsection*{Typical and periodic sets}

\begin{question}\label{q:random}
For the random sets of \cref{prop:typical}(b) with $p>\log_32$, does $T_S$ almost surely have a
divergent orbit?
\end{question}

\begin{question}\label{q:periodic}
Let $S$ be periodic modulo a power of $2$.
\begin{enumerate}[label=(\alph*)]
\item If $\pi_C(S)<\log_32$ for every closed class $C$, is $T_S$ bounded? For $S=\odd$ this
includes the statement that no Collatz orbit diverges.
\item If $\pi_C(S)>\log_32$ for some closed class, is $T_S$ unbounded? A set $S$ that is periodic
modulo a power of $2$ and misses a residue class can have provably divergent orbits: for
$S=\N\setminus(8\N-1)$ every step down is followed by two steps up, so every orbit diverges
(\cref{sec:asymmetry}). By \cref{lem:aperiodic} the residues of a divergent orbit are aperiodic.
\item Is it decidable whether $T_S$ is bounded, or whether all its orbits reach a given cycle?
Compare \cite{Con72,KS07}; uniform descent (\cref{thm:uniform}) is decidable.
\end{enumerate}
\end{question}

\begin{question}\label{q:supremum}
Is the supremum of $\pi_C(S)$ over all uniformly descending periodic sets $S$ equal to $\log_32$?
Modulo $16$ the largest value is $73/174=0.420$ (\cref{tab:uniformcount}).
\end{question}

\begin{question}\label{q:mahler}
Does $\N\setminus(4\N-1)$ contain a tail of a chain? A negative answer would show that there are
no $Z$-numbers (\cref{cor:mahler}).
\end{question}

\subsection*{The distance of the Collatz map from provable behaviour}

\begin{question}\label{q:deltak}
Does $\delta_k\to0$ (\cref{tab:deltak})? More generally, is there a set $R$ of density zero such
that every orbit of $T_S$, $S=(\odd)\triangle R$, provably reaches $(1,2)$, or at least such that
$T_S$ is provably bounded?
\end{question}

\begin{question}\label{q:addition}
Is there a set $A$ of even numbers with $\#(A\cap[1,N])=o(\log N)$ such that $T_{(\odd)\cup A}$
has a divergent orbit? \Cref{rem:heuristic} suggests not; what is the smallest constant $c$ for
which $c\log N$ added numbers below $N$ suffice?
\end{question}

\begin{question}\label{q:classes}
Find further classes of sets that are not trivially bounded and whose maps can be analysed. Others
than the periodic ones are sets defined by binary digits, Beatty sets and the primes. For the
primes, every $n\le10^7$ reaches one of the cycles $(1)$, $(2,3,5,8,4)$ and
$(31,47,71,107,161,81,41,62)$, the last one from $12.8\%$ of them.
\end{question}

\section*{Acknowledgments}
This work was carried out with the assistance of the AI models Claude Opus 5.5 and Claude
Fable 5.2 (Anthropic), which helped with the literature search, the derivations and computations,
and with the drafting of the text. The author has checked the results and takes
full responsibility for the content.

\appendix
\section{Computations}\label{sec:computations}

The ancillary files of the paper contain the C and Python scripts and the outputs of the runs (see \texttt{README.txt}).

All computations use exact integer or rational arithmetic, except where floating point is
mentioned. Floating point serves only to search (for uniformly descending sets, for the weights of
certificates, for extremal sets) and for values that are given as decimals. Every statement about
uniform descent, every certificate and every extremal set was then confirmed in exact arithmetic,
as described below. The programs are among the ancillary files,
and the file names below refer to them.

\emph{Cycles of $G$ by length} (\cref{tab:census}). For each length $\ell$ and each least element
$m$ allowed by \cref{thm:G}(c), a depth-first search over paths through vertices larger than $m$,
pruned by the fact that the lowest number reachable from $z$ in $k$ steps is $\ceil{z/2^k}$. The
run for $\ell\le30$ takes about $20$ seconds; for $\ell\le21$ the $4{,}723$ cycles were also
checked against \cref{thm:G}(c) and (d) in exact arithmetic.

\emph{Cycles of $G$ with all elements at most $N$} (\cref{q:ratio}). The numbers of such cycles are
$947$, $1{,}020$, $878{,}612$, $13{,}745{,}028$ and $1{,}133{,}265{,}572$ for $N=100,115,117,130,160$;
the jump at $N=116$ is the merger of the two nontrivial strongly connected components of the
subgraph of $G$ induced by $[1,N]$, of $54$ and $20$ vertices at $N=115$.

\emph{Short cycles} (\cref{sec:cycles}). A breadth-first search finds the shortest cycle through
$n$ inside $[1,3n]$ for all but $7{,}403$ of the $n\le10^5$. For the $10{,}001$ numbers
$n\in[9\cdot10^4,10^5]$ there is no such cycle for $742$ of them, and its length is $46$, $54$,
$62$ or $65$ for $8{,}773$ and at most $65$ for $8{,}809$ of them.

\emph{Subgraphs $\Gp\emptyset B$} (\cref{q:hereditary}). Reachable sets by depth-first search
with a cap of $10^6$ vertices; strongly connected components of the induced subgraphs by Tarjan's
algorithm and, for $B=\odd$, by intersecting the sets of vertices of $[1,2^{28}]$ reachable from
$15$ and reaching $15$ ($52{,}719{,}985$ vertices).

\emph{Primes} (\cref{q:classes}). Every orbit of $n\le10^7$ was followed until it fell below $n$
or met one of the three cycles, with a deterministic Miller--Rabin test; the largest number
reached is $106{,}780{,}274$.

\emph{Barely divergent paths} (\cref{prop:runs}(c), \cref{tab:barely}). For $(r,s)=(2,1)$ the path
from every $y<2000$ was followed for $400$ blocks with a hash set of visited numbers; repetitions
occur exactly for $y\le36$, $y\notin\{29,33,35\}$, and for $y\ge1993$ the proof of
\cref{thm:barely} applies. The thresholds $y_0=C_1/c$ of the table were computed as in that proof, in floating point,
by minimizing $|X\log3-Y\log2|$ over all $q\le q_0$, $0\le e<L$ and all starting positions in the
pattern.

\emph{Effective up-frequencies} (\cref{tab:pi,tab:deltak}). Closed classes by reachability, and
stationary distributions by Gaussian elimination over $\mathbb{Q}$ for moduli up to $64$ and in
floating point for larger moduli. The two ``straddling'' sets of \cref{tab:pi} were found by
running over all $2^{16}$ sets periodic modulo $16$ and all closed classes of their residue graphs
in floating point; every pair within $10^{-9}$ of an optimum was then recomputed over $\mathbb{Q}$.
Both optima, $39/112$ and $23/80$, are attained only by these sets (\texttt{straddle\_exact.py}).
The ``worst cycle'' of \cref{tab:deltak} is the largest mean weight of a cycle when the residues in
$S$ have weight $1$ and the others weight $0$, by Karp's algorithm with rational numbers.

\emph{Expanding cycles in integer arithmetic.} Whether a cycle of a residue graph with $a$ of its
$\ell$ vertices in $S$ is expanding, $3^a>2^\ell$, can be tested with small integers. We have
$971/1539<\log_32<665/1054$ and $665\cdot1539-971\cdot1054=1$, so no fraction with denominator at
most $2592$ lies strictly between these two fractions; hence, for $\ell\le2592$, $3^a>2^\ell$ holds
if and only if $1539a>971\ell$. Give the weight $568$ to every residue in $S$ and the weight $-971$
to every other residue. A cycle then has weight $1539a-971\ell$, and if its length is at most
$2592$, this is positive exactly when the cycle is expanding. A closed walk of positive weight
contains a simple cycle of positive weight, and the simple cycles of $\Res_k(S)$ have length at most
$2^k$. So for $k\le11$ the Bellman--Ford algorithm with these integer weights decides whether $S$
is uniformly descending, and if it is not, it produces an expanding simple cycle.

\emph{Uniform descent} (\cref{tab:uniformcount}). For all sets periodic modulo $4$, $8$ and $16$,
a first pass in floating point: the maximum mean weight of a cycle of the residue graph by Karp's
algorithm, with weights $\log\frac32$ and $-\log2$ (\texttt{unifdesc.c}). The counts and the extremal
sets of the table come from a second pass in exact arithmetic (\texttt{unifdesc\_exact.py}): uniform
descent with the integer weights above, and stationary distributions over $\mathbb{Q}$.

\emph{Smallest modifications} (\cref{tab:deltak}). Exact values by branch and bound
(\texttt{modsearch.c}): find an expanding simple cycle with the integer weights above; a uniformly
descending modification must change the step at one of its residues, and we branch on the first
residue of the cycle that is changed. For removals this finishes up to $k=7$, in about a second
for $k=7$. For arbitrary changes it finishes up to $k=6$, and for $k=7$ it shows within a few seconds that no change of fewer than $8$ residues is uniformly descending, so $\delta_7=\delta_7^-=8/128$. For $k\le6$ a plain enumeration of all changes of fewer residues confirms the lower bounds ($679{,}121$ sets for $k=6$; \texttt{bruteforce.c}). 
The same search with residues deleted from $\Res_k(\odd)$ instead of changed shows that for $k\le7$ no set of fewer than $2^k\delta_k$ residues, odd or even, meets every expanding cycle (\cref{rem:sufficient}).
For removals with $k=8$ the search for a removal of fewer than $12$ residues did not finish within $15$ minutes.
Upper bounds for $k=8,9,10$ by a randomized greedy choice in floating point (remove a random
up-vertex of an expanding cycle until none is left, then try to put back the removed residues in
random order), with $5000$, $1200$ and $1000$ trials; the sets found are confirmed by their
certificates. Lower bounds for $k=8,9,10$ by a greedy packing of pairwise disjoint expanding simple
cycles of length at most $2k+2$ (up to $22$), of which there are $2{,}831$, $12{,}481$ and
$44{,}980$; the packings consist of $10$, $14$ and $23$ cycles. The sets found
are $R_2=\{3\}$, $R_3=\{3,7\}$, $R_4=\{9,15\}$, $R_5=\{7,19,25,31\}$, $R_6=\{7,15,39,57,63\}$,
\begin{align*}
R_7={}&\{9,27,47,55,95,111,121,127\},\\
R_8={}&\{31,91,95,97,105,125,159,219,223,233,251,255\},\\
R_9={}&\{31,63,91,125,143,159,207,219,223,233,287,297,311,347,353,415,457,463,489,503,\\
&\phantom{\{}507,511\},\\
R_{10}={}&\{31,55,63,71,121,123,155,191,207,215,219,223,231,251,275,359,367,411,415,479,481,\\
&\phantom{\{}491,543,575,655,667,671,679,731,743,751,793,799,831,839,871,879,935,967,987,991,\\
&\phantom{\{}993,1019,1023\}.
\end{align*}

\emph{Certificates} (\cref{thm:mod16,tab:deltak}). For each $R_k$, in floating point: the maximum
mean cycle weight $\mu^*$ by Karp's algorithm, the potential $\psi$ of the proof of \cref{thm:uniform} for
$\mu=\mu^*/2$ (and a few other fractions of $\mu^*$), integer weights by scaling $e^\psi$ by
$10^3$ to $10^5$ and rounding. Then, exactly: a check of all edge inequalities with rational
numbers, which gives $\theta$, and a check that every $x$ with $2\le x\le B$ falls below itself (by
induction, then every such $x$ reaches $1$). The integer weights are among the ancillary files,
together with a separate program that repeats the exact part (\texttt{check\_certificate.py}). The largest stopping times in these checks are
between $2$ and $20$. The hand-made weights of \cref{tab:cert16} were found by rounding and then
adjusting by hand; the program finds slightly better ones ($\theta=0.9717$, $B=196$).

\emph{The chain of $1$} (\cref{prop:mahleradd}). The $340$ elements of $\Ch_1$ below $10^{60}$, of
which $173$ are even; below $10^3$, $10^6$, $10^{12}$ and $10^{30}$ there are $8$, $18$, $36$ and
$82$ even ones.

\emph{The heuristic} (\cref{rem:heuristic}). For each $o$ on a grid of step $1.25\cdot10^{-4}$ in
$(1/2,3/4)$, the smallest $\kappa$ with a nonnegative first exponent by bisection, and the minimum
of $\kappa/((o+\kappa)\log3-\log2)$ over the grid.

\bibliographystyle{amsplain}
\bibliography{updown}

\end{document}